\documentclass[final]{cvpr}

\makeatletter
\@namedef{ver@everyshi.sty}{}
\makeatother
\usepackage{mathptmx}
\usepackage{graphicx}
\usepackage{amsmath}
\usepackage{amsthm}
\usepackage{amssymb}

\usepackage{wrapfig}

\usepackage[pagebackref=true,breaklinks=true,colorlinks,bookmarks=false]{hyperref}

\def\cvprPaperID{372} 
\def\confYear{CVPR 2021}
\usepackage{xcolor}
\usepackage{tikz}
\usepackage{changepage}
\usepackage{tikz-3dplot}
\usepackage{tikzscale}
\usetikzlibrary{plotmarks}
\usepackage{pgfplots}
\usetikzlibrary{tikzmark,calc} 
\usetikzlibrary{decorations.pathreplacing, decorations.markings}
  
\usepackage[capbesideposition=right]{floatrow}

\newcommand*\circled[1]{\tikz[baseline=(char.base)]{
            \node[shape=circle,draw,inner sep=0.8pt] (char) {#1};}}

\xdefinecolor{navyblue}{RGB}{19 41 75}
\xdefinecolor{carolinablue}{RGB}{123 175 212}
\xdefinecolor{cb}{RGB}{123 175 212}
\xdefinecolor{flatgrey}{RGB}{162 170 173}
\xdefinecolor{matcha}{RGB}{183 224 176}

\newcommand{\dashedLayer}[6]{
			\def\a{#1} 
			\def\b{0.02}
			\def\c{#2} 
			\def\t{#3} 
			\def\d{#4} 

			\draw[line width=0.15mm, dash pattern=on 2.25pt off 1.8pt](\c+\t,0,\d) -- (\c+\t,\a,\d) -- (\t,\a,\d);                                                      
			\draw[line width=0.15mm, dash pattern=on 2.25pt off 1.8pt](\t,0,\a+\d) -- (\c+\t,0,\a+\d) node[midway,below] {#6} -- (\c+\t,\a,\a+\d) -- (\t,\a,\a+\d) -- (\t,0,\a+\d); 
			\draw[line width=0.15mm, dash pattern=on 2.25pt off 1.8pt](\c+\t,0,\d) -- (\c+\t,0,\a+\d);
			\draw[line width=0.15mm, dash pattern=on 2.25pt off 1.8pt](\c+\t,\a,\d) -- (\c+\t,\a,\a+\d);
			\draw[line width=0.15mm, dash pattern=on 2.25pt off 1.8pt](\t,\a,\d) -- (\t,\a,\a+\d);
			
			\draw[line width=0.15mm] (\c+\t,0,\d) -- (\c+\t,\a,\d);
			\draw[line width=0.15mm] (\c+\t,0,\d) -- (\c+\t,0,\a+\d);
			\draw[line width=0.15mm] (\c+\t,\a,\d) -- (\c+\t,\a,\a+\d);
			\draw[line width=0.15mm] (\c+\t,0,\a+\d) -- (\c+\t,\a,\a+\d);
			
			\filldraw[#5] (\t+\b,\b,\a+\d) -- (\c+\t-\b,\b,\a+\d) -- (\c+\t-\b,\a-\b,\a+\d) -- (\t+\b,\a-\b,\a+\d) -- (\t+\b,\b,\a+\d); 
			\filldraw[#5] (\t+\b,\a,\a-\b+\d) -- (\c+\t-\b,\a,\a-\b+\d) -- (\c+\t-\b,\a,\b+\d) -- (\t+\b,\a,\b+\d);
			\ifthenelse {\equal{#5} {}}
			{} 
			{\filldraw[#5] (\c+\t,\b,\a-\b+\d) -- (\c+\t,\b,\b+\d) -- (\c+\t,\a-\b,\b+\d) -- (\c+\t,\a-\b,\a-\b+\d);} 
		}

\newcommand{\networkLayer}[6]{
			\def\a{#1} 
			\def\b{0.02}
			\def\c{#2} 
			\def\t{#3} 
			\def\d{#4} 

			\draw[line width=0.15mm](\c+\t,0,\d) -- (\c+\t,\a,\d) -- (\t,\a,\d);                                                      
			\draw[line width=0.15mm](\t,0,\a+\d) -- (\c+\t,0,\a+\d) node[midway,below] {#6} -- (\c+\t,\a,\a+\d) -- (\t,\a,\a+\d) -- (\t,0,\a+\d); 
			\draw[line width=0.15mm](\c+\t,0,\d) -- (\c+\t,0,\a+\d);
			\draw[line width=0.15mm](\c+\t,\a,\d) -- (\c+\t,\a,\a+\d);
			\draw[line width=0.15mm](\t,\a,\d) -- (\t,\a,\a+\d);

			\filldraw[#5] (\t+\b,\b,\a+\d) -- (\c+\t-\b,\b,\a+\d) -- (\c+\t-\b,\a-\b,\a+\d) -- (\t+\b,\a-\b,\a+\d) -- (\t+\b,\b,\a+\d); 
			\filldraw[#5] (\t+\b,\a,\a-\b+\d) -- (\c+\t-\b,\a,\a-\b+\d) -- (\c+\t-\b,\a,\b+\d) -- (\t+\b,\a,\b+\d);
			\ifthenelse {\equal{#5} {}}
			{} 
			{\filldraw[#5] (\c+\t,\b,\a-\b+\d) -- (\c+\t,\b,\b+\d) -- (\c+\t,\a-\b,\b+\d) -- (\c+\t,\a-\b,\a-\b+\d);} 
		}
		
\usepackage[linesnumbered,ruled,vlined]{algorithm2e}

\SetCommentSty{mycommfont}
\SetKwInput{Input}{Input}                
\SetKwInput{Output}{Output}         
\SetKwInput{Dataset}{Dataset}     
\SetKwInput{Settings}{Settings}      
\SetKwInput{Initialization}{Initialization}      
\SetAlgorithmName{Alg.}{Alg.}{List of Algorithms} 
\makeatletter
\newcommand{\algorithmfootnote}[2][\footnotesize]{%
  \let\old@algocf@finish\@algocf@finish
  \def\@algocf@finish{\old@algocf@finish
    \leavevmode\rlap{\begin{minipage}{\linewidth}
    #1#2
    \end{minipage}}%
  }%
}
\makeatother

\usepackage{floatrow}

\usepackage{caption} 
\usepackage{subcaption}
\usepackage{makecell}
\usepackage{booktabs}
\usepackage{multirow}

\newcolumntype{P}[1]{>{\centering\arraybackslash}p{#1}}

\usepackage{hyperref} 
  
\newtheoremstyle{bfnote}%
  {}{}
  {\itshape}{}
  {\bfseries}{.}
  { }{\thmname{#1}\thmnumber{ #2}\thmnote{ (#3)}}
\theoremstyle{bfnote}
\newtheorem{theorem}{Theorem}

\newtheorem*{notation*}{Notation}

\newtheorem{app_theorem}{Theorem}[section]
\newtheorem{app_corollary}{Corollary}[section]

\newtheorem{app_definition}{Definition}[section]

\usepackage{url}

\newcommand\blfootnote[1]{%
  \begingroup
  \renewcommand\thefootnote{}\footnote{#1}%
  \addtocounter{footnote}{-1}%
  \endgroup
}

\definecolor{olive}{rgb}{0.42, 0.56, 0.14}

\begin{document}

\title{Discovering Hidden Physics Behind Transport Dynamics}

\author{Peirong Liu\textsuperscript{1} \quad Lin Tian\textsuperscript{1} \quad Yubo Zhang\textsuperscript{1} \quad Stephen Aylward\textsuperscript{3} \quad Yueh Lee\textsuperscript{2} \quad Marc Niethammer\textsuperscript{1} \vspace{0.3cm} \\  
\textsuperscript{1}Department of Computer Science, University of North Carolina at Chapel Hill, Chapel Hill, USA\\ \textsuperscript{2}Department of Radiology, University of North Carolina at Chapel Hill, Chapel Hill, USA\\ 
\textsuperscript{3}Kitware Inc., New York, USA \\ \vspace{0.1cm}
{\tt\small \{peirong, lintian, zhangyb, mn\}@cs.unc.edu} \quad {\tt\small stephen.aylward@kitware.com} \quad {\tt\small yueh\_lee@med.unc.edu}
}

\maketitle


\begin{abstract}
Transport processes are ubiquitous. They are, for example, at the heart of optical flow approaches; or of perfusion imaging, where blood transport is assessed, most commonly by injecting a tracer. An advection-diffusion equation is widely used to describe these transport phenomena. Our goal is estimating the underlying physics of advection-diffusion equations, expressed as velocity and diffusion tensor fields. We propose a learning framework (\texttt{YETI}) building on an auto-encoder structure between 2D and 3D image time-series, which incorporates the advection-diffusion model. To help with identifiability, we develop an advection-diffusion simulator which allows pre-training of our model by supervised learning using the velocity and diffusion tensor fields. Instead of directly learning these velocity and diffusion tensor fields, we introduce representations that assure incompressible flow and symmetric positive semi-definite diffusion fields and demonstrate the additional benefits of these representations on improving estimation accuracy. We further use transfer learning to apply \texttt{YETI} on a public brain magnetic resonance (MR) perfusion dataset of stroke patients and show its ability to successfully distinguish stroke lesions from normal brain regions via the estimated velocity and diffusion tensor fields.
\blfootnote{This work was supported by the National Institutes of Health (NIH) under award number NIH 2R42NS086295.}
\end{abstract}

\section{Introduction}
\label{sec: intro}  

Many transport phenomena can be formalized by partial differential equations (PDEs). However, numerically solving PDEs is expensive for high spatial dimensions and across timescales~\cite{bar2019pde}. Significant developments in deep learning have recently led to an explosive growth of data-driven solutions for PDEs via deep neural networks (DNNs). This work either directly models the solution via DNNs~\cite{weinan2018ritz,pinn2019,smith2020eikonet} or learns mesh-free, infinite-dimensional operators with DNNs \cite{lu2020deeponet,bhattacharya2020reduction,nelsen2020random,sitzmann2020implicit,li2020fourier}. 

Despite of the significant progress achieved by the aforementioned methods for \emph{solving PDEs} forward, \emph{inverse} PDE problems, i.e., parameter estimation, remains challenging~\cite{lieberman2010invpde,tartakovsky2020pinn}. Such problems have been extensively explored in the context of optical flow and for general image registration, where the underlying PDE model is typically an advection equation and the sought-for parameter is a displacement or velocity vector field~\cite{horn1981determining,borzi2003optimal,modersitzki2004_numerical,beg2005computing,hart2009optimal}. DNN solutions have also been studied~\cite{flownet,yang2017quicksilver,balakrishnan2019voxelmorph,shen2019networks}.

In contrast, we are interested in estimating the spatially-varying velocity \emph{and} diffusion \emph{tensor} fields (termed physics parameter fields in what follows) for more general advection-diffusion equations. Challenges arise from identifiability (i.e., if observed transport is due to advection or diffusion), physical plausibility (e.g., for fluid flow, vector fields should be divergence-free) and diffusion tensor structure (i.e., predominant direction and anisotropy). Further, while optical flow and registration approaches typically deal with image pairs, the parameter estimation for advection-diffusion equations is generally based on time-series data.

Limited work to estimate the parameter fields of advection-diffusion equations exist. Tartakovsky et al.~\cite{tartakovsky2020pinn} do not consider advection or diffusion tensors, but instead proposed a DNN to learn 2D diffusion fields only from diffusion PDEs. B\'ezenac et al.~\cite{emmanuel2018} learn 2D velocity and diffusion fields of advection-diffusion PDEs by DNNs. Liu et al.~\cite{liu2020piano} proposed an optimization approach to estimate velocity and diffusion fields of an advection-diffusion equation in 3D. Though promising, the numerical optimization approach is time-consuming, especially when dealing with large datasets. Koundal et al.~\cite{koundal2020omt} use optimal mass transport combined with a spatially constant diffusion. Note that all aforementioned methods assume isotropic diffusion (i.e., not general diffusion tensors), which is insufficient to accurately model complex materials (e.g., anisotropic porous media, brain tissue) where diffusion is mostly anisotropic. Further, both the approaches by B\'ezenac et al.~\cite{emmanuel2018} and Liu et al.~\cite{liu2020piano} suffer from potential identifiability issues as they are both only based on fitting the observed time-series data without explicitly controlling how to allocate between the velocity and diffusion fields; though an inductive bias is introduced by considering divergence-free vector fields only, via a loss term in~\cite{emmanuel2018} and a special parameterization in~\cite{liu2020piano}.
 
We therefore introduce a novel learning framework, termed \texttt{YETI}, which works in 2D and 3D, to discover the underlying velocity and diffusion tensor fields from observed advection-diffusion time-series. 
Our contributions are:
\\ \vspace{-0.6cm}
\begin{itemize}
\item {\it A novel advection-diffusion learning model}. Given an advection-diffusion process, \texttt{YETI} not only predicts the transport dynamics, but also reconstructs the underlying velocity and diffusion tensor fields.\\ \vspace{-0.65cm}
\item {\it Representation theorems for divergence-free vectors and symmetric PSD tensors}. Our estimates are grounded in theorems ensuring realistic constraints on the learned physics parameter fields \emph{by construction}. This also helps to significantly improve the diffusion reconstruction by providing supervision on its anisotropic structure during training. \\ \vspace{-0.65cm}
  \item {\it Advection-diffusion simulator.} We develop a simulator for quasi-realistic advection-diffusion that can be used for physics-parameter-supervised model training.\\ \vspace{-0.65cm}
\item {\it A two-phase learning strategy}. 
(1) We initially train our model on a simulated dataset with ground truth physics to improve identifiability when estimating velocity and diffusion from advection-diffusion; 
(2) We apply PDE-based transfer learning for transport dynamics via a time-series auto-encoder, which allows us to discover unknown velocity and diffusion fields of advection-diffusion processes for real world data.  

\end{itemize}

\section{Background and Problem Setup}
\label{sec: setup}
Advection-diffusion PDEs are used to describe a large family of physical processes, e.g., fluid dynamics, heat conduction, and wind dynamics~\cite{emmanuel2018}. Advection refers to the transport with fluid flow, diffusion is driven by the gradient of mass concentration. 
Let $C({\mathbf{x}}, t)$ denote the mass concentration at location ${\mathbf{x}}$ in a bounded domain $\Omega\subset \mathbb{R}^d\,(d = 2,\,3)$, at time $t$. The local mass concentration changes of an advection-diffusion process are described by the PDE
\vspace{-0.15cm}
\begin{equation}
\frac{\partial C({\mathbf{x}}, t)}{\partial t} = \underbrace{-  \nabla \left({\mathbf{V}}({\mathbf{x}}) \cdot C({\mathbf{x}}, t) \right)}_{\text{Fluid flow}} + \underbrace{\nabla \cdot \left({\mathbf{D}}({\mathbf{x}})\, \nabla C({\mathbf{x}}, t)\right)}_{\text{Diffusion}},
\label{eq: full_adv_diff} 
\vspace{-0.15 cm}
\end{equation}
for specified boundary conditions (B.C.). The spatially-varying velocity field ${\mathbf{V}}$ (${\mathbf{V}}(\mathbf{x}) \in \mathbb{R}^d$) and diffusion tensor field ${\mathbf{D}}$ (${\mathbf{D}}(\mathbf{x}) \in \mathbb{R}^{d\times d}$) describe the advection and diffusion. 

\vspace{-0.25 cm}
\paragraph{Incompressible Flow} 
In fluid mechanics, incompressibility describes a flow with constant density within a parcel of fluid moving with the flow velocity. This is a common assumption for fluids in practice as the density variation is negligible in most scenarios \cite{kim19deepfluid}. Mathematically, the velocity field of an incompressible flow has zero divergence (i.e., is divergence-free), which simplifies Eq.~(\ref{eq: full_adv_diff}) to
\vspace{-0.15cm}
\begin{equation}
\frac{\partial C({\mathbf{x}}, t)}{\partial t} = \underbrace{- {\mathbf{V}}({\mathbf{x}})\cdot\nabla C({\mathbf{x}}, t)}_{\text{Incompressible flow}} + \underbrace{\nabla \cdot \left({\mathbf{D}}({\mathbf{x}})\, \nabla C({\mathbf{x}}, t)\right)}_{\text{Diffusion}}. 
\label{eq: adv_diff}   
\vspace{-0.15 cm}
\end{equation}
Note when $\mathbf{D} \to 0$, Eq.~(\ref{eq: adv_diff}) is simply an advection equation, which is the basis for many variational optical flow or image registration methods~\cite{horn1981determining,borzi2003optimal,modersitzki2004_numerical,beg2005computing,hart2009optimal,deepflow2013,flownet,yang2017quicksilver,balakrishnan2019voxelmorph,shen2019networks}. 

\vspace{-0.25 cm}
\paragraph{Symmetric Positive Semi-definite (PSD) Diffusion}
Diffusion can be generally modeled via symmetric positive-definite tensors\footnote{For simplicity $\mathbf{D}$ is often chosen as a scalar field or even as a constant, which amounts to an isotropic diffusion assumption. However, diffusion in complex materials (e.g., anisotropic porous media, brain tissue) is generally anisotropic, hence requiring the estimation of general PSD tensors.}. In practice, diffusion tensors, $\mathbf{D}$, are assumed to be symmetric positive semi-definite (PSD)~ \cite{niethammer2006dti}. 

\vspace{-0.25 cm}
\paragraph{Perfusion Imaging}
\label{sec: perfusion} 
Using an intravascular tracer, perfusion imaging is used to quantify blood flow through brain parenchyma~\cite{demeestere2020stroke}. Its resulting tracer concentration image time-series can be viewed as the tracer being transported by blood flow within the vessels (advection) while diffusing within the extracellular space (diffusion)~\cite{strouthos2010dilution,harabis2013dilution,cookson2014spatial,liu2020piano}. Perfusion imaging is the motivating application behind our approach, yet our approach applies generally to parameter estimation for advection-diffusion equations.  

\section{Constraint-free Representations}
\label{sec: repre}
As explained in Sec.~\ref{sec: setup}, incompressibility and symmetric PSD-ness are commonly used realistic assumptions for fluid flow and diffusion. This section introduces two theorems to represent divergence-free velocity fields and symmetric PSD diffusion tensor fields \emph{based on potentials and representative matrices} to obtain divergence-free vector fields and PSD tensors \emph{by construction}, even at test time.

\subsection{Divergence-free Vector Representation}
\label{sec: div_free}
\vspace{-0.1cm}
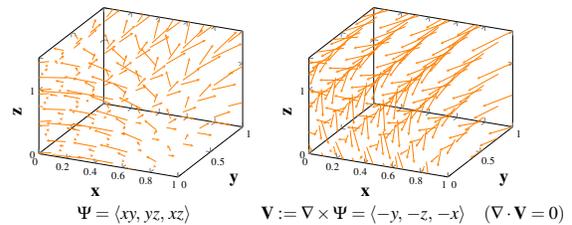
\begin{figure}[h]
\vspace*{-0.3cm} 
\centering
\resizebox{0.9\textwidth}{!}{
\begin{tikzpicture}
  \begin{axis}[
    domain=-1:1,
    samples=10,
    xmin=0,xmax=1,
    ymin=0,ymax=1,
    zmin=0,zmax=1.5,
    xlabel={\LARGE$\mathbf{x}$},
    ylabel={\LARGE$\mathbf{y}$},
    zlabel={\LARGE$\mathbf{z}$},
    ]
    \pgfplotsinvokeforeach{0, .5, 1, 1.5, 2}{
      \addplot3[quiver,-stealth,
      quiver={
      u = x * y,	
      v = y * z,
      w = x * z,
        colored = orange,
        scale arrows=.175
        }]
      (x,y,#1);
    }
  \end{axis}
    \node at (3.2, -1.25){\LARGE $\boldsymbol{\Psi} =  \langle xy,\, yz, \, xz \rangle$};
  \end{tikzpicture}
  
\begin{tikzpicture}
  \begin{axis}[
    domain=-1:1,
    samples=10,
    xmin=0,xmax=1,
    ymin=0,ymax=1,
    zmin=0,zmax=1.5,
    xlabel={\LARGE$\mathbf{x}$},
    ylabel={\LARGE$\mathbf{y}$},
    zlabel={\LARGE$\mathbf{z}$},
    ]
    \pgfplotsinvokeforeach{0, .5, 1, 1.5, 2}{
      \addplot3[quiver,-stealth,
      quiver={
      u = - y,	
      v = -  z,
      w = -  x,
        colored = orange,
        scale arrows=.175
        }]
      (x,y,#1);
    }
  \end{axis}
    \node at (3.5, -1.25){$\LARGE \mathbf{V} := \nabla \times \boldsymbol{\Psi} =  \langle -y, \, -z, \, -x \rangle \quad (\nabla \cdot {\mathbf{V}} =  0)$};

  \end{tikzpicture}
  }
  \caption{Representing a 3D divergence-free vector field ($\mathbf{V}$) by the curl of vector potentials ($\boldsymbol{\Psi}$) via Eq. (\ref{eq: repre_v}).}
  \label{fig: div_free}
 \vspace{-0.3cm}
\end{figure}

B\'ezenac et al.~\cite{emmanuel2018} penalize deviations of the learned velocity fields from zero divergence. However, zero divergence cannot be guaranteed and is only encouraged during training. Kim et al.~\cite{kim19deepfluid} parametrize velocity vectors via the curl of vector fields, but do not consider the boundary conditions that should be imposed on the vector potentials in bounded domain scenarios~\cite{dubois1990div_free,amrouche1998div_free,maria2003hhd,amrouche2013div_free}. 
Therefore, we seek a representation which enables us to learn velocity fields $\mathbf{V}$ on a domain $\Omega\subset \mathbb{R}^d\,(d = 2,\,3)$ with smooth boundary such that 
(1) $\mathbf{V}$ is divergence-free {\it{by construction}}; 
and (2) {\it{any}} divergence-free $\mathbf{V}$ can be represented. 

Specifically, we introduce the representation theorem for divergence-free velocity fields below. (See complete proof for Theorem~\ref{thm: repre_v} in Supp. \ref{app: proof_div_free}.)  \vspace{-0.2cm}
\begin{theorem}[Divergence-free Vector Field Representation by the Curl of Potentials]For any vector field $\mathbf{V} \in L^p(\Omega)^d$ on a bounded domain $\Omega\subset\mathbb{R}^d$ with smooth boundary $\partial \Omega$. If $\mathbf{V}$ satisfies $\nabla \cdot \mathbf{V} = 0$, there exists a potential $\boldsymbol{\Psi}$ in $L^p(\Omega)^{\alpha}$ such that ($\alpha = 1$ when $d = 2$, $\alpha = 3$ when $d = 3$)
\vspace{-0.15cm}
\begin{equation}
\mathbf{V} = \nabla \times \boldsymbol{\Psi}, \quad \boldsymbol{\Psi} \cdot \mathbf{n}\big|_{\partial \Omega} = 0,\, \boldsymbol{\Psi} \in L^{p}(\Omega)^{\alpha}.
\label{eq: repre_v}
\vspace{-0.15cm}
\end{equation}
Conversely, for any $\boldsymbol{\Psi}\in L^p(\Omega)^{\alpha}$, $\nabla \cdot \mathbf{V} =\nabla \cdot (\nabla \times \boldsymbol{\Psi})= 0$. 
\label{thm: repre_v}
\end{theorem}
\vspace{-0.2cm}
With Theorem~\ref{thm: repre_v}, we learn a divergence-free velocity field via its potential $\boldsymbol{\Psi}$, to ensure the divergence-free property of ${\mathbf{V}}$ while bypassing direct constraints on ${\mathbf{V}}$ itself.

\subsection{Symmetric PSD Tensor Representation}
\label{sec: psd}
\vspace{-0.1cm}
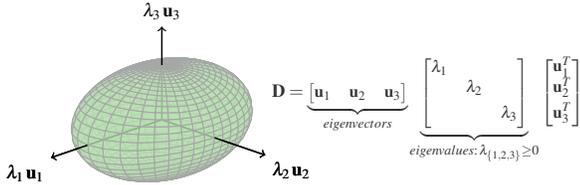
\begin{figure}[h]
\vspace*{-0.4cm}
\hspace*{-0.3cm}
\centering
\resizebox{1\textwidth}{!}{
\newcommand{\asa}{1}
\newcommand{\bsa}{0.5}
\newcommand{\csa}{0.3}
\tdplotsetmaincoords{70}{135}
\begin{tikzpicture}[scale=2,tdplot_main_coords,line join=bevel,fill opacity=.8]
    \pgfsetlinewidth{.1pt}
    \tdplotsphericalsurfaceplot{72}{36}
        {1/sqrt((sin(\tdplottheta))^2*(cos(\tdplotphi))^2/\asa+
        (sin(\tdplottheta))^2*(sin(\tdplotphi))^2/\bsa + (cos(\tdplottheta))^2/\csa)} 
       {gray!70} 
         {matcha} 
        {\draw[color=black,thick,->] (0,0,0) -- (1.5,0,0) node[anchor=north east] {$\lambda_1 \, \mathbf{u}_1$};}
        {\draw[color=black,thick,->] (0,0,0) -- (0,1.4,0) node[anchor=north west]{$\lambda_2 \, \mathbf{u}_2$};}
        {\draw[color=black,thick,->] (0,0,0) -- (0,0,0.95) node[anchor=south]{$\lambda_3 \, \mathbf{u}_3$};}
        
        \node at (-1.8, 1.8, 0.1) {$
{\small {\mathbf{D}}= \underbrace{
	\begin{bmatrix}
	\mathbf{u}_1 &  \mathbf{u}_2 & \mathbf{u}_3 
	\end{bmatrix}}_{eigenvectors}
	\underbrace{
	\begin{bmatrix}
	\lambda_1 &  & \\ \vspace{1mm} 
	 & \lambda_2 &  \\ \vspace{1mm} 
	 & & \lambda_3
	\end{bmatrix}}_{eigenvalues:\, \lambda_{\{1,2,3\}}\geq 0}
	 \begin{bmatrix} 
	 \mathbf{u}_1^T \\ \vspace{1mm} \mathbf{u}_2^T \\ \vspace{1mm}  \mathbf{u}_3^T 
	 \end{bmatrix}}
$};
\end{tikzpicture}
	}
  \caption{Representing a 3D symmetric PSD diffusion tensor ($\mathbf{D}$) by eigenvectors ($\mathbf{U}$) and eigenvalues ($\boldsymbol{\Lambda}$) via Eq. (\ref{eq: spectral}).}
  \label{fig: psd}
\end{figure}

\vspace{-0.1cm}
Similar to Sec.~\ref{sec: div_free}, we aim for a representation for tensors $\mathbf{D}$ such that  
(1) $\mathbf{D}$ is a symmetric PSD tensor {\it{by construction}}, and 
(2) {\it any} symmetric PSD $\mathbf{D}$ can be represented.  
\vspace{-0.15cm}
\begin{notation*}
\vspace{-0.15cm}
Denote the $n\times n$ symmetric PSD tensor group as
\vspace{-0.15cm}
\begin{equation}
PSD(n) \equiv \{ \mathbf{D} \in \mathbb{R}^{n\times n}\,\big|\, \forall \mathbf{x}\in\mathbb{R}^n: \mathbf{x}^T\mathbf{D} \mathbf{x} \geq 0 \}.
\label{eq: psd_group}
\vspace{-0.15cm}
\end{equation}
\end{notation*}

First, consider the spectral decomposition of $\mathbf{D}$:  
\vspace{-0.15cm}
\begin{equation}
\mathbf{D} = \mathbf{U}\boldsymbol{\Lambda} \mathbf{U}^{T}, \quad {\mathbf{D}} \in PSD(n),\, \mathbf{U} \in SO(n),\, \boldsymbol{\Lambda}  \in SD(n),
\label{eq: spectral}
\vspace{-0.15cm}
\end{equation}
where the columns of $\mathbf{U}$ are the eigenvectors of $\mathbf{D}$, which belong to the special real orthogonal group $SO(n)$:
\vspace{-0.15cm}
\begin{equation}
SO(n) \equiv \{  \mathbf{U} \in \mathbb{R}^{n \times n}\, \big| \, \mathbf{U}^T\mathbf{U} = \mathbf{I}, \, \text{det}(\mathbf{U}) = 1 \},
\vspace{-0.15cm}
\end{equation}
and $\boldsymbol{\Lambda}$ are the corresponding non-negative eigenvalues, in the special group of non-negative diagonal matrices:
\vspace{-0.15cm}
\begin{equation}
SD(n) \equiv \{diag(\lambda_1,\,...\, \lambda_n) \in \mathbb{R}^{n \times n} \,\big| \, \lambda_1,\,...,\, \lambda_n \geq 0\}.
\vspace{-0.15cm}
\end{equation} 
Combining the surjective Lie exponential mapping on $SO(n)$ ($exp:\, \mathfrak{so}(n) \mapsto  SO(n)$, $\mathfrak{so}(n)$ is the group of skew-symmetric matrices)~\cite{lezcano2019spectral} with the isomorphic mapping from the space of upper triangular matrices with zero diagonal entries to $\mathfrak{so}(n)$ ($\alpha:\, \mathbb{R}^{\frac{n(n-1)}{2}} \mapsto \mathfrak{so}(n)$)~\cite{lezcano2019trivializations}, we give the following representation theorem for symmetric PSD tensors. (See complete proof for Theorem~\ref{thm: repre_psd} in Supp.~\ref{app: proof_psd}). \vspace{-0.2cm}
\begin{theorem}[Symmetric PSD Tensor Representation by Spectral Decomposition]
For any tensor $\mathbf{D}  \in PSD(n)$, there exists an upper triangular matrix with zero diagonal entries, $\mathbf{B} \in \mathbb{R}^{\frac{n(n-1)}{2}}$, and a diagonal matrix with non-negative diagonal entries, $\boldsymbol{\Lambda} \in SD(n)$, satisfying:
\vspace{-0.15cm}
\begin{equation}
\mathbf{D} = \mathbf{U}\, \boldsymbol{\Lambda} \, \mathbf{U}^T,\quad \mathbf{U} = exp(\mathbf{B} - \mathbf{B}^T) \in SO(n).
\label{eq: repre_psd}
\vspace{-0.15cm}
\end{equation}
Conversely, for any upper triangular matrix with zero diagonal entries, $\mathbf{B} \in \mathbb{R}^{\frac{n(n-1)}{2}}$, and any diagonal matrix with non-negative diagonal entries, $\boldsymbol{\Lambda} \in SD(n)$, Eq.~(\ref{eq: repre_psd}) results in a symmetric PSD tensor, $\mathbf{D}\in PSD(n)$. 
\label{thm: repre_psd}
\end{theorem}
\vspace{-0.2cm}
Therefore, we can learn a symmetric PSD diffusion tensor via its representative matrices $\mathbf{B}$ and $\boldsymbol{\Lambda}$ based on Theorem~\ref{thm: repre_psd}, to ensure the symmetric PSD property of ${\mathbf{D}}$ while not imposing it on the learning space. 
In implementation, we use the Cayley retraction \cite{lezcano2019trivializations} to approximate the exponential mapping on $SO(n)$, to reduce computational costs.

\section{\texttt{YETI}: discovering hidden phYsics bEhind Transport dynamIcs}
\label{sec: network}


\input{sub/network/fig/fw} 


Sec.~\ref{sec: repre} described the representation theorems for divergence-free vector fields and symmetric PSD tensor fields, which allow imposing constraints by construction. This section introduces our learning scheme, \texttt{YETI}, for discovering the divergence-free velocity fields ($\mathbf{V}$) and symmetric PSD diffusion fields ($\mathbf{D}$) underlying an observed advection-diffusion process. \texttt{YETI} consists of two phases: (1) \emph{Direct physics learning}: reconstructs physics parameters ($\mathbf{V},\, \mathbf{D}$) from input time-series of transport dynamics, trained on a simulated dataset under the supervision of ground truth physics parameters; (2) \emph{Latent physics learning}: transfers the pre-trained model from the direct physics learning phase to real mass concentration time-series. As the ground truth physics parameters are unknown in this case, transfer learning proceeds via a time-series auto-encoder which integrates the advection-diffusion PDE.

Due to the large size of the concentration time-series, especially for 3D domains, \texttt{YETI} trains on patches. Given a time-series $C = \{ C^{t_i} \in \mathbb{R}(\Omega) \big| i = 1,\, 2,\, ...,\, N_{T} \}$, we randomly extract $32^3$ ($32^2$ for 2D domains) patches from same spatial locations ($\Omega_p\subset\Omega$), across $N_{\text{in}}$ time points. Each training sample ($C_p$) starts from a randomly selected time point $t_i$ ($i \in \{1,\, 2,\,...,\, N_{T}-N_{\text{in}}+1\}$), resulting in $C_p = \{ C_p^{t_j} \in \mathbb{R}(\Omega_p) \big| j = i,\, ...,\, i+N_{\text{in}}-1 \}$ (Fig.~\ref{fw} (top left)).

\subsection{Direct Physics Learning Phase}
\label{sec: vd_net}

\texttt{YETI} can use different convolutional neural network architectures. Here we modify the networks from \cite{3dunet2016} (\cite{2dunet2015}) to estimate $\mathbf{V}$, $\mathbf{D}$ on 3D (2D) domains. As shown in Fig.~\ref{fw} (bottom left), $C_p$ is first processed by a \emph{dynamics} encoder which extracts latent features from input time-series. The images of the input time-series are simply treated as input channels. 
Next, two separate decoders learn the mappings to the potentials of $\mathbf{V}$ and $\mathbf{D}$. 
The $\mathbf{V}$-decoder outputs the potential $\widehat{\boldsymbol{\Psi}}$ to represent the corresponding divergence-free velocity field $\widehat{\mathbf{V}}$ via Theorem~\ref{thm: repre_v}. Likewise, the $\mathbf{D}$-decoder outputs $\widehat{\mathbf{B}}$ concatenated with $\widehat{\boldsymbol{\Lambda}}$ to express the reconstructed symmetric PSD diffusion tensor field $\widehat{\mathbf{D}}$ via Theorem~\ref{thm: repre_psd}. The reconstruction loss of the predicted $\widehat{\mathbf{V}},\, \widehat{\mathbf{D}}$ is 
\vspace{-0.2cm}
\begin{equation}
\mathcal{L}_{\mathbf{VD}} = \frac{1}{|\Omega_p|}\int_{\Omega_p} \big\| \mathbf{V}- \widehat{\mathbf{V}} \big\|_{2} + \big\| \mathbf{D} - \widehat{\mathbf{D}} \big\|_{\text{F}}\, d\mathbf{x}\,,
\label{eq: l1-vd}
\vspace{-0.2cm}
\end{equation}
where $\mathbf{V},\, \mathbf{D}$ denote the ground truth physics parameters and $\|\cdot\|_{2}$, $\|\cdot\|_{\text{F}}$ the vector 2-norm and tensor Frobenius norm.
\vspace{-0.2cm}
\paragraph{Structure-informed Supervision}
\label{sec: structure_informed}
In order to improve the network's ability to capture the anisotropic structure of diffusion tensors, we further impose supervision on the eigenvectors ($\widehat{\mathbf{U}}$) and eigenvalues ($\widehat{\boldsymbol{\Lambda}}$) of $\widehat{\mathbf{D}}$ (Fig.~\ref{fig: psd}). Note $\widehat{\mathbf{U}} = [\widehat{\mathbf{u}}_1,\, \widehat{\mathbf{u}}_2(,\, \widehat{\mathbf{u}}_3)]$ is an intermediate output from $\widehat{\mathbf{B}}$ via Eq.~(\ref{eq: repre_psd}).
\vspace{-0.2cm}
\begin{equation}
\hspace{-0.2cm}
\mathcal{L}_{\mathbf{U}\boldsymbol{\Lambda}} = \frac{1}{|\Omega_p|}\int_{\Omega_p} \sum_{i=1}^{3(2)}min\big\{\big\|\mathbf{u}_i \pm \widehat{\mathbf{u}}_i \big\|_{2}\big\}  + \big\| \boldsymbol{\Lambda} - \widehat{\boldsymbol{\Lambda}} \big\|_{\text{F}}\,d\mathbf{x}\,,
\label{eq: l1-str}
\vspace{-0.2cm}
\end{equation}
where the element-wise $min$ is used for resolving the sign ambiguities regarding the eigenvector directions.

Overall, the complete loss for direct physics learning is
\vspace{-0.15cm}
 \begin{equation}
\mathcal{L}_{\text{Dir}} = \mathcal{L}_{\mathbf{VD}} + w_{\mathbf{U}\boldsymbol{\Lambda}} \, \mathcal{L}_{\mathbf{U}\boldsymbol{\Lambda}}\,,\quad w_{\mathbf{U}\boldsymbol{\Lambda}} > 0\,. 
\label{eq: l1}
\end{equation}
 



\subsection{Latent Physics Learning Phase}
\label{sec: integration}

In this phase, we transfer the pre-trained model from Sec.~\ref{sec: vd_net} to time-series datasets where the ground truth ${\mathbf{V}}, \, {\mathbf{D}}$ are unknown, thus training is supervised by the transport dynamics.  
Specifically, we use an advection-diffusion PDE solver to integrate the initial state ($C_{p}^{t_i}$) forward in time to $t_{i+N_{\text{out}}-1}$ via Eq.~(\ref{eq: adv_diff}) (Fig.~\ref{fw} (bottom right)), and fine-tune the physics reconstruction model by minimizing the differences between the predicted ($\widehat{C}_p$) and the input ($C_p$) time-series. 

 
\vspace{-0.3cm}
\paragraph{Numerical Solution} 
\label{sec: numerical}
We use a first-order upwind scheme \cite{leveque2002} to approximate the differential operators of the advection term in Eq.~(\ref{eq: adv_diff}), and a nested central-forward-backward difference scheme for the diffusion term. Discretizing all the spatial derivatives on the right side of Eq.~(\ref{eq: adv_diff}) results in a system of ordinary differential equations (ODEs), which can be solved by numerical integration. We use the RK45 method to advance in time ($\delta t$) to predict $\widehat{C}^{t+\delta t}$. Note when the input mass transport time-series has relatively large temporal resolution ($\Delta t$), the chosen $\delta t$ should be smaller than $\Delta t$ (Fig.~\ref{fw} (top right)) to satisfy the Courant-Friedrichs-Lewy (CFL) condition \cite{gottlieb2000ssp,leveque2002}, thereby ensuring stable numerical integration. (See numerical discretization details and stability discussions in Supp. \ref{app: numericals}.)

 
\vspace{-0.3cm}
\paragraph{Patch-based Boundary Conditions}
\label{sec: bc} 
Improperly-chosen boundary conditions (B.C.) can cause stability issues during numerical integration. As the boundaries ($\partial \Omega_p$) of extracted patches are typically not the actual domain boundaries ($\partial \Omega$), they need to be carefully specified. We set up Cauchy B.C. \cite{tartakovsky2020pinn} as virtual B.C. for training patches: 
\vspace{-0.2cm}
\begin{equation}
\underbrace{\widehat{C}_p^{t_j}\big|_{\partial \Omega_p} = C_p^{t_j}\big|_{\partial \Omega_p}}_{\text{\circled{1} Dirichlet}}, ~~ \underbrace{\frac{\partial \widehat{C}_p^{t_j}}{\partial \mathbf{n}}\bigg|_{\partial \Omega_p} = 0}_{\vspace{-0.1cm}\text{\circled{2} Zero-Neumann}}\,, ~~ (j = i,\, ..., \, i+N_{\text{out}}-1)  
\label{eq: virtual_bc} 
\vspace{-0.2cm}
\end{equation}
$C_p^{t_j}, \, \widehat{C}_p^{t_j}$ denote the input and predicted mass concentration for patch $C_p$ at time $t_j$, respectively. $\mathbf{n}$ is the outward unit vector normal to $\partial \Omega_p$. 
By imposing Eq.~(\ref{eq: virtual_bc}) we assume no flux across $\partial \Omega_p$ (zero-Neumann B.C.) and values on $\partial \Omega_p$ reflect the actual amount of mass (Dirichlet B.C.). 
We further discard the boundaries of predicted time-series ($\widehat{C}_p$) to reduce potential artifacts introduced by the virtual B.C.

\vspace{-0.2cm}
\paragraph{Losses} 

Given an input training sample $\{C_p^{t_j} \in \mathbb{R}(\Omega_p)| j = i,\, \ldots,\, i+N_{\text{in}}-1\}$, we compute the mean squared error of the predicted time series $\widehat{C}_p$ at output collocation time points $t_i,\, \ldots,\, t_{i+N_{\text{out}}-1}$, to encourage predictions close to observed values. We also account for spatial gradient differences \cite{mathieu2016gdl} at each collocation time point. Therefore, the collocation concentration loss ($\mathcal{L}_{\text{CC}}$) is defined as
\vspace{-0.15 cm} 
\begin{equation} 
\mathcal{L}_{\text{CC}} = \frac{1}{N_{\text{out}}}\hspace{-0.1cm} \sum_{j = i}^{i+N_{\text{out}}-1} \hspace{-0.15cm}\int_{\Omega_p} \hspace{-0.15cm} \frac{\big| C_p^{t_j} - \widehat{C}_p^{t_j} \big|^2 + w_{\nabla}\, \big\| \nabla C_p^{t_j} - \nabla \widehat{C}_p^{t_j} \big\|_2^2}{\vert\Omega_p\vert} \, d{\bf{x}},
\label{eq: loss_cc}
\vspace{-0.15 cm} 
\end{equation} 
where $\nabla$ denotes the spatial derivative, $w_{\nabla} > 0$.


Assuming the physics parameter fields are spatially smooth, we add a regularization term ($\mathcal{L}_{\text{SS}}$) on the gradient fields of each component of the reconstructed $\widehat{\mathbf{V}},\, \widehat{\mathbf{D}}$:
\vspace{-0.2 cm} 
\begin{equation}
 \mathcal{L}_{\text{SS}} = \frac{1}{\vert\Omega_p\vert} \int_{\Omega_p} \big (\sum_{i = 1}^{3(2)}\, \| \nabla \widehat{\mathbf{V}}_i \|_2^2 + \sum_{i = 1}^{9(4)}\, \| \nabla \widehat{\mathbf{D}}_i \|_{2}^2 \big )\,d{\bf{x}}\,. 
 \label{eq: ss}
\vspace{-0.15 cm} 
\end{equation} 

Overall, the complete loss for latent physics learning is
\vspace{-0.15 cm} 
\begin{equation}
\mathcal{L}_{\text{Lat}} = \mathcal{L}_{\text{CC}} + w_{\text{SS}}\, \mathcal{L}_{\text{SS}}\,,\quad w_{\text{SS}} > 0\,. 
\label{eq: l2} 
\end{equation}


\section{Experimental Results}
\label{sec: exp}
 
In Sec.~\ref{exp: 2d_demo}-\ref{exp: ixi}, we test on simulated time-series in 2D and 3D. We show the significant improvements achieved by \texttt{YETI}'s direct physics learning with structure-informed supervision. In Sec.~\ref{exp: isles}, we transfer \texttt{YETI}'s pre-trained model on simulated data to real time-series of MR perfusion images from stroke patients. We demonstrate \texttt{YETI}'s ability in distinguishing stroke lesions from normal brain regions via its reconstructed physics parameters fields ($\widehat{\mathbf{V}}$, $\widehat{\mathbf{D}}$). 

\subsection{2D Simulation: Anisotropic Moving Gaussian}
\label{exp: 2d_demo}

\input{sub/exp/fig/2d_demo_plt} 

\begin{figure*}[t]
	\noindent\resizebox{1\textwidth}{!}{
	\begin{tikzpicture} 
	
	\node at (-6, -1) {Err(C)};
	
	\node at (-6, -2) {\includegraphics[width=0.175\textwidth]{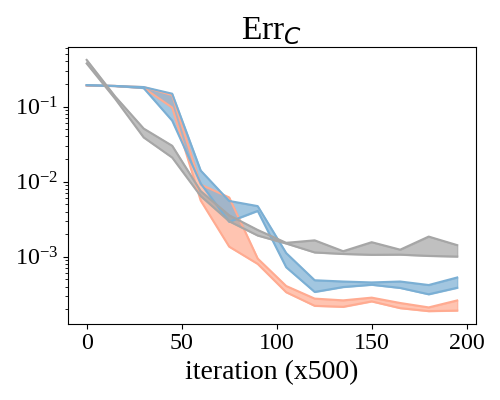}};
	\node at (-3, -2) {\includegraphics[width=0.175\textwidth]{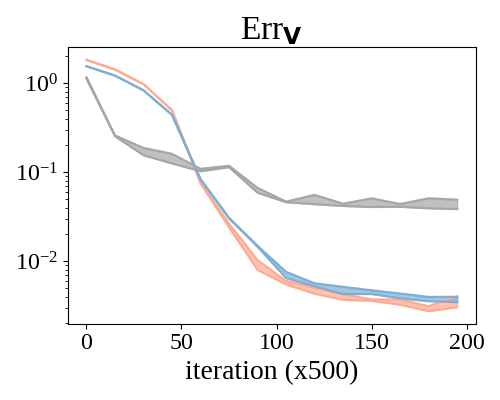}};
	\node at (0, -2) {\includegraphics[width=0.175\textwidth]{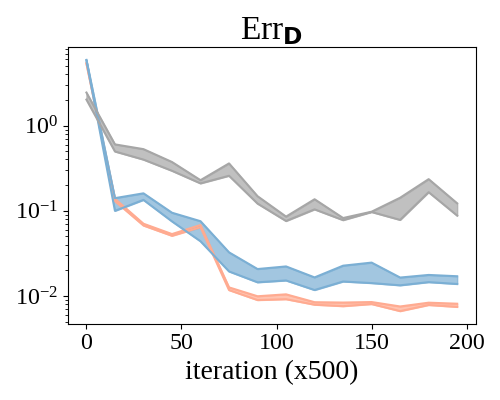}};
	\node at (3, -2) {\includegraphics[width=0.175\textwidth]{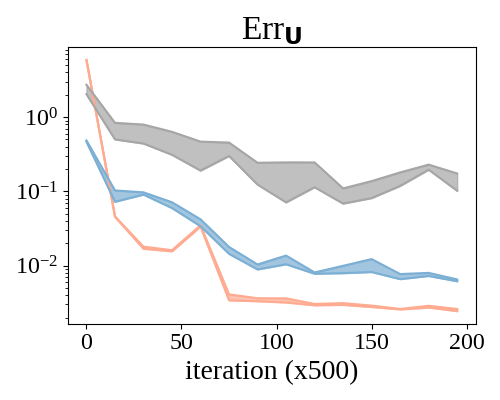}};
	\node at (6, -2) {\includegraphics[width=0.175\textwidth]{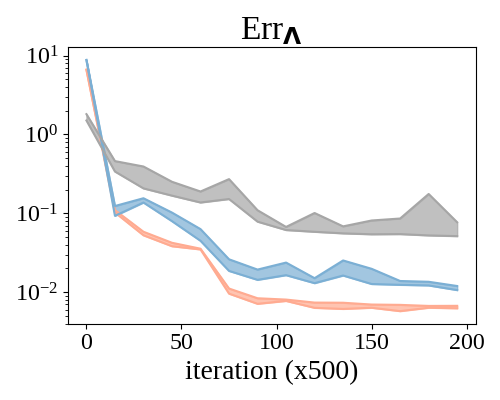}};

	\end{tikzpicture}
	}
	\vspace{-0.7cm} 
	\caption{Mean relative absolute error (RAE) of ``dynamics-supervised'' \texttt{YETI} (grey), ``$\mathbf{VD}$-supervised'' \texttt{YETI} (blue) and ``structure-informed'' \texttt{YETI} (orange) for anisotropic moving Gaussian. Horizontal axes indicate training iterations, vertical axes show RAE in log scale. The banded curves indicate the 25\% \& 75\% percentile of the errors among 50 test samples.}
	\label{fig: 2d_demo_loss_comp}
\end{figure*}

\paragraph{Dataset}
\label{sec: 2d_dataset} 
We simulate advection-diffusion in 2D on-the-fly (Fig.~\ref{fig: 2d_demo}). Each sample is a 2D image time-series of size $64^2 \times 40$ (on a $64^2$ domain with $1\,mm$ uniform spacing; $N_T = 40$ with time interval $\Delta t = 0.01\,s$). Every time-series records the advection-diffusion of mass concentration initialized by a Gaussian ($\sigma = 2$) at a randomly selected center and transported by ${\mathbf{V}},\, {\mathbf{D}}$, computed by potential $\boldsymbol{\Psi},\, \mathbf{B},\, \boldsymbol{\Lambda}$. Specifically, the nonzero entries in $\mathbf{B}$ are assigned values randomly, resulting in random diffusion eigenvectors directions. All components of $\boldsymbol{\Lambda}$ are randomly sampled from range $[0, 1]$, and $\boldsymbol{\Psi}$ is randomly sampled within $[-10, 10]$, to assure numerical stability of the simulation (See Supp.~\ref{app: numericals}). 

\vspace{-0.2cm}
\paragraph{Experiments}
\label{sec: 2d_exp} 
We compare three learning settings: (1) ``Dynamics-supervised'' \texttt{YETI}, where we only supervise on the time-series and directly train in the latent physics learning phase (line \hyperlink{code: 7}{7}-\hyperlink{code: 14}{14} (Alg.~\hyperlink{alg}{1})). (2) ``$\mathbf{VD}$-supervised'' \texttt{YETI}, where we train in the direct physics learning phase (line \hyperlink{code: 1}{1}-\hyperlink{code: 6}{6} (Alg.~\hyperlink{alg}{1})) yet without structure-informed supervision. I.e., $\mathcal{L}_{\mathbf{VD}}$ is the only training loss. (3) ``Structure-informed'' \texttt{YETI}, the proposed direct physics learning phase (Sec.~\ref{sec: vd_net}). Specifically, the input time-series length for all models is $N_{\text{in}} = 10$. For ``dynamics-supervised'' \texttt{YETI}, we set $N_{\text{out}} = 10$, $w_{\nabla} = 0.5,\, w_{\text{SS}} = 0.1$. For ``structure-informed'' \texttt{YETI}, we set $w_{\mathbf{U}\boldsymbol{\Lambda}} = 0.5$. We use \texttt{Adam} with learning rate $10^{-3}$ and a decay factor of 0.1 per 500 iterations, for all models. We test on 50 samples per 500 training iterations. Reconstructed physics ($\widehat{\mathbf{V}},\, \widehat{\mathbf{D}},\, \widehat{\mathbf{U}},\, \widehat{\boldsymbol{\Lambda}}$) on the entire domain are obtained by splicing the output patches together. By solving the advection-diffusion PDE forward with $\widehat{\mathbf{V}},\, \widehat{\mathbf{D}}$ we obtain the predicted time-series $\widehat{C}$ on the original domain.   

For evaluation, we compute the following mean relative absolute error (RAE) for the reconstructed $\widehat{\mathbf{V}}, \,\widehat{\mathbf{D}},\,\widehat{\mathbf{U}},\, \widehat{\boldsymbol{\Lambda}}$: 
\vspace{-0.18 cm}
\begin{equation}
Err_{\mathbf{F}} = \frac{1}{|\Omega|}\int_{\Omega} \| \mathbf{F} - \widehat{\mathbf{F}} \|/\| \mathbf{F} \| \, d\mathbf{x}\,,
\label{eq: eval}
\vspace{-0.18 cm}
\end{equation}
where $\mathbf{F},\, \widehat{\mathbf{F}}$ denote ground truth and prediction, $\|\cdot\|$ is the absolute, 2-norm or Frobenius norm for scalars, vectors or tensors. Time-series error ($Err_{C}$) is the average mean RAE over time-series predicted on all collocation time points. 

Fig.~\ref{fig: 2d_demo_loss_comp} compares the reconstruction errors of the three models throughout learning. Eventually all models achieve comparable accuracy with respect to the time-series prediction. However, without explicit supervision on $\mathbf{V}$ and $\mathbf{D}$, ``dynamics-supervised'' \texttt{YETI} leads to much higher errors on reconstructed $\widehat{\mathbf{V}},\, \widehat{\mathbf{{D}}}$. While ``$\mathbf{VD}$-supervised'' \texttt{YETI} reaches similar performance to ``structure-informed'' \texttt{YETI} with respect to $\widehat{\mathbf{V}}$, it results in higher $Err_{\mathbf{D}}$, which can also be observed from its worse performance in reconstructing the structural diffusion quantities ($Err_{\mathbf{U}},\, Err_{\boldsymbol{\Lambda}}$). This performance gap becomes much larger in Sec.~\ref{exp: ixi}, for more complex advection-diffusion patterns in 3D. 

\input{sub/exp/fig/vd_generator} 

\subsection{3D Simulation: Brain Advection-Diffusion}
\label{exp: ixi}
\input{sub/exp/fig/ixi_plt}

\begin{figure*}[t]
	\noindent\resizebox{1\textwidth}{!}{
	\begin{tikzpicture} 
	
	\node at (-6, -2) {\includegraphics[width=0.175\textwidth]{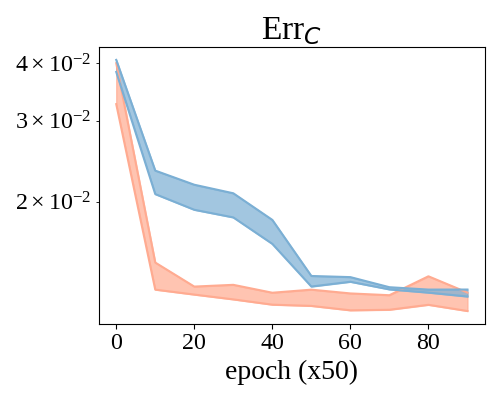}};
	\node at (-3, -2) {\includegraphics[width=0.175\textwidth]{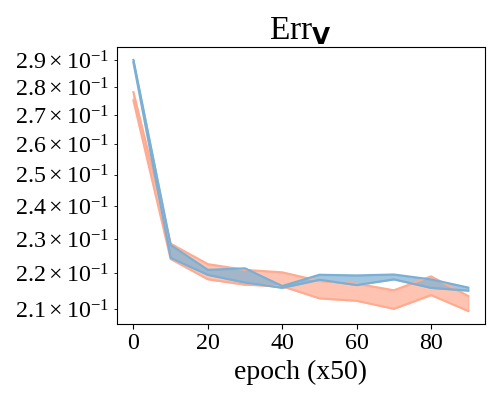}};
	\node at (0, -2) {\includegraphics[width=0.175\textwidth]{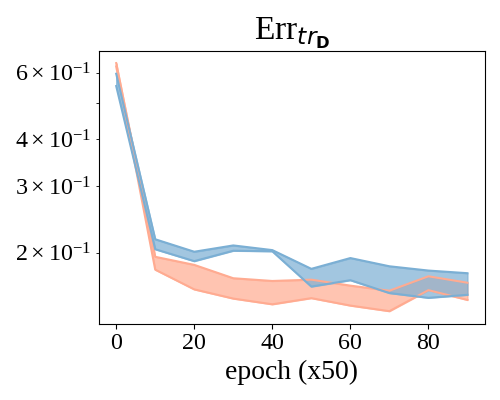}};
	\node at (3, -2) {\includegraphics[width=0.175\textwidth]{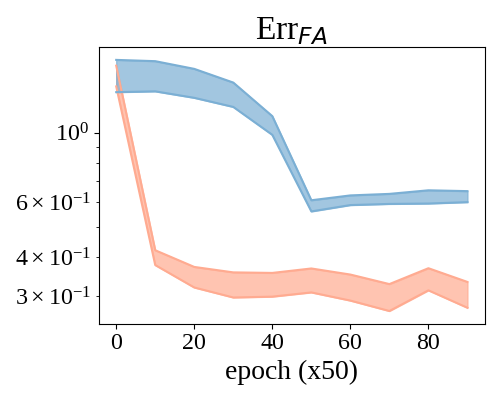}};
	\node at (6, -2) {\includegraphics[width=0.175\textwidth]{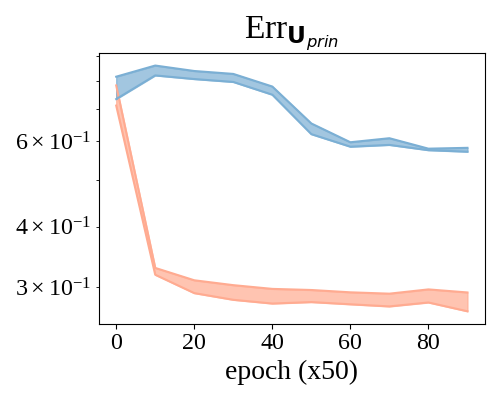}};

	\end{tikzpicture}
	}
	\vspace{-0.7cm} 
	\caption{Mean relative absolute error (RAE) of ``$\mathbf{VD}$-supervised'' \texttt{YETI} (blue) and ``structure-informed'' \texttt{YETI} (orange) for brain advection-diffusion dataset. Horizontal axes indicate training epoch, vertical axes show RAE in log scale. The banded curves indicate the 25\% \& 75\% percentile of the errors among 40 test samples.} 
	\label{fig: ixi_loss_comp}
\end{figure*}

\paragraph{Dataset}
\label{sec: ixi_data}

We developed a brain advection-diffusion simulator based on the \texttt{IXI} brain dataset\footnote{Available for download: \url{http://brain-development.org/ixi-dataset/}.}. We use 200 patients with complete collections of T1-/T2-weighted images, MRA, and diffusion-weighted images (DWI) 
to simulate 3D velocity and diffusion tensor fields\footnote{Our goal is to obtain nontrivial velocity and diffusion tensor fields for 3D advection-diffusion simulation. These will likely not be realistic perfusion simulations, but are beneficial for network pre-training, similar in spirit to the synthetic datasets for optical flow training~\cite{flownet,mayer2016large}.}. All images are resampled to isotropic spacing ($1\,mm$) and rigidly registered intra-subject by \texttt{ITK}~\cite{itk}. We simulate velocity fields from brain vessels segmented by \texttt{ITK-TubeTK} using T1-/T2-weighted and MRA images (Fig.~\ref{fig: vd_generator} (top)). Diffusion tensors are estimated from the DWIs using \texttt{Dipy}~\cite{dipy} (Fig.~\ref{fig: vd_generator} (middle)). We scale the value range of velocity fields to $[-1,\, 1]$ and that of diffusion fields to $[-0.2,\, 0.2]$. Accounting for various transport effects from advection and diffusion, we additionally compute velocity and diffusion fields with $50\%$ of the original velocity and diffusion values. For each brain advection-diffusion sample, the initial concentration state is assumed to be given by the MRA image with intensity ranges rescaled to $[0,\,1]$. Time-series (length $N_T = 40$, interval $\Delta t = 0.1\,s$) are then simulated given these velocity and diffusion tensor fields by our advection-diffusion PDE solver (Fig.~\ref{fig: vd_generator} (bottom)). Thus the simulated dataset includes 800 brain advection-diffusion time-series (4 time-series for each of the 200 subjects). We randomly select 40 time-series for validation and testing, respectively. (Supp.~\ref{app: ixi} describes the simulation in detail.) 

\vspace{-0.2cm}
\paragraph{Experiments}
\label{sec: ixi_exp}

We test the same three models described in Sec.~\ref{sec: 2d_exp} with input time-series length $N_{\text{in}} = 5$ for all models. For ``dynamics-supervised'' \texttt{YETI}, we set $N_{\text{out}} = 5$, $w_{\nabla} = 1,\, w_{\text{SS}} = 0.1$. For ``structure-informed'' \texttt{YETI}, we set $w_{\mathbf{U}\boldsymbol{\Lambda}} = 1$. All models are trained using \texttt{Adam} with learning rate $5\times10^{-4}$ and a decay factor of 0.1 every 50 epochs. 

For comparison, we compute $\| \mathbf{V} \|_2$, the $2$-norm map for the reconstructed velocity fields. To analyze the reconstructed diffusion tensors, we consider three diffusion scalar maps widely used in diffusion tensor imaging \cite{mukherjee2008dti}: (1) Trace ($tr_{\mathbf{D}}$), sum of tensor eigenvalues ($\boldsymbol{\Lambda}$), indicating the overall diffusion strength; (2) Fractional anisotropy (FA),
\vspace{-0.15 cm}
\begin{equation}
\text{FA} =  \sqrt{\frac{1}{2}}\sqrt{\frac{(\lambda_1 - \lambda_2)^2 + (\lambda_2 - \lambda_3)^2 + (\lambda_3 - \lambda_1)^2}{\lambda_1^2 + \lambda_2^2 + \lambda_3^2}},
\vspace{-0.15 cm}
\label{eq: fa}
\end{equation} 
where $\boldsymbol{\Lambda} = diag\big({\lambda_1,\, \lambda_2,\, \lambda_3}\big)$, measuring the amount of diffusion anisotropy; (3) 
Principal diffusion orientation ($\mathbf{U}_{\text{prin}}$), the eigenvector corresponding to the largest eigenvalue. The element-wise absolute value of $\mathbf{U}_{\text{prin}}$ scaled by FA is used for color-by-orientation (CbO) visualization. Mean RAE (Eq.~(\ref{eq: eval})) is used to evaluate the reconstruction performances on $\widehat{C}$, $\widehat{\mathbf{V}}$, and $tr_{\mathbf{D}}$, FA, $\mathbf{U}_{\text{prin}}$ of $\widehat{\mathbf{D}}$.


Fig.~\ref{fig: ixi} visualizes the reconstruction results of the three models. Without supervision on the physics parameter fields, ``dynamics-supervised'' \texttt{YETI} cannot identify the velocity and diffusion tensor fields well from the advection-diffusion time-series: the reconstructed $\widehat{\mathbf{V}}$ is mixed up with $\widehat{\mathbf{D}}$, resulting in a noisy $\|\widehat{\mathbf{V}}\|_2$ map and $\widehat{\mathbf{D}}$ loses most local structure. While ``$\mathbf{VD}$-supervised'' \texttt{YETI} achieves much better reconstruction performance regarding the overall magnitudes of velocity and diffusion (i.e., $\|\mathbf{V}\|_2,\, tr_{\mathbf{D}}$), it struggles to infer the anisotropic diffusion structure, which results in unrealistic FA and CbO maps. In contrast, the proposed ``structure-informed'' \texttt{YETI} achieves significant reconstruction improvements and, in particular, successfully captures diffusion anisotropy. Fig.~\ref{fig: ixi_loss_comp} also illustrates this effect by comparing the mean RAE of ``$\mathbf{VD}$-supervised'' and ``structure-informed'' \texttt{YETI} on all test samples. Although the two models eventually achieve similar $Err_C$, $Err_{\mathbf{V}}$ and $Err_{tr_{\mathbf{D}}}$, without guidance on the diffusion tensor structure (by their eigenvectors and eigenvalues), $\mathbf{VD}$-supervised'' \texttt{YETI} tends to get stuck in sub-optimal local minima and thus does not learn the anisotropic diffusion structure well.

\subsection{ISLES2017: Brain Perfusion Dataset}
\label{exp: isles}

Perfusion images (Sec.~\ref{sec: perfusion}) reflect local changes of injected tracer transport across time, and are widely used to assess cerebrovascular diseases, including acute stroke~\cite{demeestere2020stroke}. For stroke patients, different observed tracer concentration time-series between the lesion and normal regions can indicate abnormal perfusion patterns, e.g., insufficient blood flow to a particular region of the brain.

\vspace{-0.2cm}
\paragraph{Dataset}
\label{sec: isles_data}

We test on the public Ischemic Stroke Lesion Segmentation (\texttt{ISLES}) 2017 dataset~\cite{isles2015a}, including 75 (43 training, 32 testing) ischemic stroke patients. Each patient has a 4D dynamic susceptibility contrast (DSC) MR perfusion image (with 40 to 80 available time points, time interval $\approx 1\,s$)~\cite{essig2013mrp}. Gold-standard lesion segmentations are provided for patients in the training set. All images are resampled to isotropic spacing (1mm) and rigidly registered intra-subject via \texttt{ITK}. Based on the relation between MR signal and tracer concentration~\cite{fieselmann2011sig2ctc}, we obtain a tracer concentration time-series for each patient, $\{C^{t_i} \in \mathbb{R}(\Omega) \vert\, i = 1,\, 2,\, \ldots,\, N_T\}$, where $t_1$ is the time-to-peak for total concentration over the entire brain, at which we assume the injected tracer has been fully transported into the brain \cite{liu2020piano}. We randomly select 10 patients with lesion maps for testing. The remaining 65 concentration time-series are flipped along the axial axis for data augmentation, resulting in a total of 130 time-series samples, from which we randomly select 10 samples for validation, the others are for training.

\input{sub/exp/fig/isles_plt}
\begin{table}[t]
\resizebox{\linewidth}{!}{
\centering
    \begin{tabular}{P{0.2cm}P{0.5cm}P{0.7cm}P{0.6cm}P{0.6cm}P{0.6cm}P{0.6cm}P{0.6cm}P{0.6cm}P{0.6cm}} 
       \toprule \\[-3ex] 
      \multicolumn{2}{c}{\multirow{2}{*}{\bf Metrics}} & \multicolumn{3}{c}{\bf \texttt{YETI}} & \multicolumn{2}{c}{\bf \texttt{PIANO}}  &  \multicolumn{3}{c}{\bf \texttt{ISLES}}  \\ [-0.7ex]
        \cmidrule(lr){3-5}
        \cmidrule(lr){6-7}
        \cmidrule(lr){8-10}
       & & ${\| {\bf{V}} \|_2}$ & $tr_{\mathbf{D}}$ & ${\mathbf{U}_{\text{prin}}}$ & ${\| {\bf{V}} \|_2}$ & $D$ & CBF & CBV & MTT \\ [-0.2ex]
     \midrule\\[-3ex]
     
      \multirow{3}{*}{\thead{$\mu^r$\\($\downarrow$)}} & {\emph{Me.}} & {\bf 0.37} & 0.81 & - & 0.55 & 0.61 & 0.68 & 0.78 & 0.59 \\ [-0.4ex]
       &  {\emph{Med.}} & {\bf 0.35} & 0.79 & - & 0.56 & 0.58 & 0.69 & 0.82 & 0.61 \\ 
       &  {\emph{STD}} & {\bf 0.11} & 0.17 & - & 0.13 & 0.18 & 0.19 & 0.34 & 0.20 \\ 
     \hline\\[-2.3ex]
      \multirow{3}{*}{\thead{$|t|$\\($\uparrow$)}} & {\emph{Me.}} & {\bf 111.33} & 36.28 & - & 67.76 & 32.86 & 39.26 & 15.98 & 31.83 \\ 
       &  {\emph{Med.}} & {\bf 130.38} & 43.28 & - & 54.16 & 38.93 & 29.09 & 8.48 & 27.67 \\ 
       &  {\emph{STD}} & {\bf 69.56} & 17.02 & - & 49.16 & 20.47 & 36.38 & 17.79 & 31.61 \\ 
     \hline\\[-2.4ex]
      \multirow{3}{*}{\thead{$\angle$\\($\uparrow$)}} &  {\emph{Me.}}  & - & - & 54.13$^\circ$ & - & - & - & - & -  \\ 
       &  {\emph{Med.}} & - & - & 54.99$^\circ$  & - & - & - & - & - \\ 
       &  {\emph{STD}} & - & - & 5.96$^\circ$  & - & - & - & - & -  \\ [-0.5ex] 
       
\midrule[\heavyrulewidth] \\ [-3.5ex]
\multicolumn{8}{l}{\footnotesize$^*$ $\downarrow$ ($\uparrow$) indicates the lower (higher) values are better.} \\ [-0.5ex]
\bottomrule  \\ [-3.2ex] 
    \end{tabular}  
    \caption{Quantitative comparison between \texttt{YETI}, \texttt{PIANO} and \texttt{ISLES} maps across 10 test subjects, using \emph{Mean (Me.)}, \emph{Median (Med.)}, \emph{Standard Deviation (STD)} of relative mean $\mu^r$ and mean principal diffusion angle deviation $\angle$.} 
    \label{tab: isles}
}

\end{table}
 

\vspace{-0.2cm}
\paragraph{Experiments}
\label{sec: isles_exp}

We transfer the pre-trained ``structure-informed'' \texttt{YETI} (Sec.~\ref{sec: ixi_exp}) on the \texttt{ISLES} tracer concentration time-series data (Sec.~\ref{sec: integration}, line \hyperlink{code: 7}{7}-\hyperlink{code: 14}{14} (Alg.~\hyperlink{alg}{1})). Specifically, we set $N_{\text{in}} = N_{\text{out}} = 5$, $w_{\nabla} = 0.5,\, w_{\text{SS}} = 0.1$. We use the \texttt{Adam} optimizer with learning rate set to $10^{-4}$.

As in the \texttt{PIANO} approach proposed by Liu et al.~\cite{liu2020piano}, we compute two feature maps for the reconstructed velocity: (1) ${\mathbf{V}}_{\text{rgb}}$: color-coded orientation map of $\widehat{\mathbf{V}}$; (2) $\| {\mathbf{V}} \|_2$: $2$-norm of $\widehat{\mathbf{V}}$. \texttt{PIANO} models diffusion as scalar fields ($D$), which is directly used as a feature map. We use trace ($tr_{\mathbf{D}}$) and color-by-orientation (CbO) maps, introduced in Sec.~\ref{sec: ixi_exp}, as \texttt{YETI}'s feature maps for its reconstructed diffusion tensor fields. Note the $tr_{\mathbf{D}}$ map reflects overall diffusion magnitudes, which is similar to the $D$ map in \texttt{PIANO}.

Fig.~\ref{fig: isles_plt} shows \texttt{YETI}'s feature maps for four test patients; all are highly consistent with the lesion regions. Details of the blood flow trajectories are revealed in ${\mathbf{V}}_{\text{rgb}}$ by the ridged patterns and the sharp color changes in the unaffected hemisphere, while the flat patterns within the stroke lesion provide little directional information about the velocity. From $\|{\mathbf{V}} \|_2$ and $tr_{\mathbf{D}}$, one can easily locate the lesion where the magnitudes are low. The CbO maps also show abnormalities in lesions, where the lower FA (darker) and the mottled colors reveal inconsistent $\mathbf{U}_{\text{prin}}$ with little anisotropic diffusion pattern. Note that \texttt{YETI}'s reconstruction depends on the \emph{observed} advection-diffusion, i.e., it will not capture effects in regions where the tracer was never transported to. 

\vspace{-0.2cm}
\paragraph{Comparisons} 
We compare \texttt{YETI}'s feature maps with (1) \texttt{PIANO}~\cite{liu2020piano} feature maps ($\|\mathbf{V}\|_2,\, D$) and (2) \texttt{ISLES}~\cite{isles2015a} perfusion summary maps (Cerebral blood flow (CBF), Cerebral blood volume (CBV), Mean transit time (MTT)). Specifically, we focus on the differences between lesions and normal regions revealed by the features maps of the above methods. We compute three metrics between lesions and their contralateral regions (c-lesion) obtained by mirroring lesions to the unaffected side via the midline of the cerebral hemispheres: 
(1) Relative mean ($\mu^r \in [0, 1]$): 
\vspace{-0.15 cm}
\begin{equation} 
\mu^r = min\bigg\{\frac{\text{mean in lesion}}{\text{mean in c-lesion}}, \frac{\text{mean in c-lesion}}{\text{mean in lesion}}\bigg\},
\label{eq: rel_mean}
\vspace{-0.15 cm}
\end{equation} 
where $min$ accounts for typically larger MTT (while other metrics are typically smaller) in lesion than c-lesion; (2) Absolute t-value ($|t|$): the absolute value of the unpaired t-statistic between lesion and c-lesion; (3) Mean principal diffusion angle deviation ($\angle$): the average angle between $\mathbf{U}_{\text{prin}}$ in lesion and the mirrored angle in c-lesion ($\mathbf{U}_{\text{prin}}^c$). The direction ambiguity of eigenvectors is resolved by taking $\angle = min\big\{\angle(\pm\mathbf{U}_{\text{prin}}, \,\mathbf{U}_{\text{prin}}^c)\big\}$. Note that $\angle$ is specific to \texttt{YETI} as \texttt{PIANO} does not estimate diffusion \emph{tensors}.

Table~\ref{tab: isles} compares maps from \texttt{YETI}, \texttt{PIANO} and \texttt{ISLES} based on the above three metrics for the 10 test patients. \texttt{YETI}'s $\| {\mathbf{V}} \|_2$ maps achieve much lower $\mu^r$ with respect to either mean, median or standard deviation, indicating significantly smaller velocity magnitudes in the stroke lesions compared to normal c-lesion regions. This can also be seen in the absolute t-values ($|t|$) of $\| {\mathbf{V}} \|_2$ which are much larger than for all other metrics from \texttt{PIANO} and \texttt{ISLES}. Furthermore, \texttt{YETI} provides insights on the deviations of the principal diffusion orientations in the lesions, where the mean deviation angle is around $54^{\circ}$ with a standard deviation of $5.96^{\circ}$. Overall, \texttt{YETI}'s measures show more sensitivity in distinguishing stroke from normal regions than the measures of the competing \texttt{PIANO} approach and the standard perfusion measures provided as part of \texttt{ISLES}.

\section{Conclusions}
\label{sec: con}
We introduced \texttt{YETI}, a learning framework to estimate the divergence-free velocity vector fields and symmetric PSD diffusion tensor fields underlying observed advection-diffusion processes. \texttt{YETI} is pre-trained with simulated datasets, which helps improve identifiability with respect to the estimated advection and diffusion. Simulation experiments in 2D and 3D demonstrate \texttt{YETI}'s ability to resolve velocity and diffusion ambiguities and to recover diffusion anisotropy. Further, we used transfer learning via a time-series auto-encoder formulation to apply \texttt{YETI} on real brain perfusion data of stroke patients. Experiments show that \texttt{YETI} successfully detects abnormalities of velocity and diffusion tensor fields in stroke lesions and provides more sensitive measures than competing approaches.




{\small
\bibliographystyle{ieee_fullname}
\bibliography{piano_cvpr}
}


\clearpage
\onecolumn
\appendix

\renewcommand\thefigure{\thesection.\arabic{figure}}

{\centering 
\section*{Discovering Hidden Physics Behind Transport Dynamics\\Supplementary Material}
}
\vspace{0.55cm}
This supplementary material contains proofs for our representation theorems and additional implementation details for \texttt{YETI}. Supp.~\ref{app: proof_div_free} and Supp.~\ref{app: proof_psd} give the proofs for Theorem~\ref{thm: repre_v} and Theorem~\ref{thm: repre_psd}, respectively. Supp.~\ref{app: numericals} introduces our advection-diffusion PDE toolkit in \texttt{PyTorch} and discusses numerical discretization and stability conditions. Supp.~\ref{app: ixi} provides detailed descriptions of our brain advection-diffusion simulation dataset, including how we construct the velocity and diffusion fields, and how we simulate the brain advection-diffusion time-series. 

\vspace{0.7cm}


\section{Theorem 1: Divergence-free Vector Representation by the Curl of Potentials}
\label{app: proof_div_free}

{\it
\noindent For any vector field $\mathbf{V} \in L^p(\Omega)^d$ on a bounded domain $\Omega\subset\mathbb{R}^d$ with smooth boundary $\partial \Omega$. If $\mathbf{V}$ satisfies $\nabla \cdot \mathbf{V} = 0$, there exists a potential $\boldsymbol{\Psi}$ in $L^p(\Omega)^{\alpha}$ such that ($\alpha = 1$ when $d = 2$, $\alpha = 3$ when $d = 3$)
\begin{equation}
\mathbf{V} = \nabla \times \boldsymbol{\Psi}, \quad \boldsymbol{\Psi} \cdot \mathbf{n}\big|_{\partial \Omega} = 0,\, \boldsymbol{\Psi} \in L^{p}(\Omega)^{\alpha}.
\label{app_eq: repre_v}
\end{equation}
Conversely, for any $\boldsymbol{\Psi}\in L^p(\Omega)^{\alpha}$, $\nabla \cdot \mathbf{V} =\nabla \cdot (\nabla \times \boldsymbol{\Psi})= 0$.  
}

{\noindent  (Here, $L^{p}$ refers to the space of measurable functions for which the $p$-th power of the function absolute value is Lebesgue integrable. Specifically, let $1 \leq p < \infty$ and $(\Omega, \Sigma, \mu)$ be a measure space. $L^p(\Omega)$ space is the set of all measurable functions whose absolute value raised to the $p$-th power has a finite integral, i.e.,
$\|f\|_p \equiv \left( \int_{\Omega} |f|^p\;\mathrm{d}\mu \right)^{1/p}<\infty$.)}

\begin{proof}
We give a brief proof in this supplementary material to make it self-contained. Please refer to \cite{dubois1990div_free,amrouche1998div_free,maria2003hhd,amrouche2013div_free} for additional discussions regarding alternative boundary conditions for the potentials and domains with complex geometry.

\begin{app_definition}[The space of curl of potentials]
\begin{equation}
\mathcal{H}_{\text{curl}}(\Omega) \equiv \{  \nabla \times \boldsymbol{\Psi}  \, \big| \, \boldsymbol{\Psi} \in L^p(\Omega)^{\alpha},\,\Omega \in \mathbb{R}^d \}, 
\end{equation}
where $\alpha = 1$ when $d = 2$, $\alpha = 3$ when $d = 3$. 
\end{app_definition}

\begin{app_definition}[The space of divergence-free velocity fields]
\begin{equation}
\mathcal{H}_{\text{div}}(\Omega) \equiv \{  \mathbf{V} \in L^p(\Omega)^{d} \, \big| \,  \nabla \cdot \mathbf{V} = 0,\, \Omega \in \mathbb{R}^d \}, 
\end{equation}
where $\alpha = 1$ when $d = 2$, $\alpha = 3$ when $d = 3$. 
\end{app_definition}

\begin{itemize}
\item Clearly, for $\forall\, \mathbf{V}\in \mathcal{H}_{\text{curl}}(\Omega)$, $\nabla \cdot \mathbf{V} = 0$.

Specifically, when $d=3$, the curl of a vector field $\boldsymbol{\Psi} = \Psi_x\mathbf{i} + \Psi_y\mathbf{j} + \Psi_z\mathbf{k}$ is computed by
\begin{align}
\nonumber \nabla \times \boldsymbol{\Psi} &=
\begin{vmatrix} 
\mathbf{i} & \mathbf{j} & \mathbf{k} \\[5pt]
{\dfrac{\partial}{\partial x}} & {\dfrac{\partial}{\partial y}} & {\dfrac{\partial}{\partial z}} \\[10pt]
\Psi_x & \Psi_y & \Psi_z 
\end{vmatrix} = \left(\frac{\partial \Psi_z}{\partial y} - \frac{\partial \Psi_y}{\partial z}\right) \mathbf{i} + \left(\frac{\partial \Psi_x}{\partial z} - \frac{\partial \Psi_z}{\partial x} \right) \mathbf{j} + \left(\frac{\partial \Psi_y}{\partial x} - \frac{\partial \Psi_x}{\partial y} \right) \mathbf{k} \nonumber\\[1ex] 
&= \bigg[
\frac{\partial \Psi_z}{\partial y} - \frac{\partial \Psi_y}{\partial z}, \,
\frac{\partial \Psi_x}{\partial z} - \frac{\partial \Psi_z}{\partial x}, \,
\frac{\partial \Psi_y}{\partial x} - \frac{\partial \Psi_x}{\partial y}
\bigg]^T. 
\label{eq: curl3D}
\end{align}
Therefore, 
\begin{equation}
\nabla \cdot (\nabla \times \Psi) = \frac{\partial}{\partial x} \bigg(\frac{\partial \Psi_z}{\partial y} - \frac{\partial \Psi_y}{\partial z}\bigg) 
+ \frac{\partial}{\partial y} \bigg(\frac{\partial \Psi_x}{\partial z} - \frac{\partial \Psi_z}{\partial x}\bigg) 
+ \frac{\partial}{\partial z} \bigg(\frac{\partial \Psi_y}{\partial x} - \frac{\partial \Psi_x}{\partial y}\bigg) = 0.
\end{equation}

When $d=2$, we can express $\Psi$ as $ \Psi = \Psi_x\mathbf{i} + \Psi_y\mathbf{j}\,(+\,0\,\mathbf{k})$. Likewise,
\begin{equation}
\nabla \times\Psi=
\begin{vmatrix} 
\mathbf{i} & \mathbf{j} & \mathbf{k} \\[5pt]
{\dfrac{\partial}{\partial x}} & {\dfrac{\partial}{\partial y}} & {\dfrac{\partial}{\partial z}} \\[10pt]
\Psi_x & \Psi_y & 0 
\end{vmatrix}
= \bigg[
0, \, 0, \, \frac{\partial \Psi_y}{\partial x} - \frac{\partial \Psi_x}{\partial y}
\bigg]^T,
\label{eq: curl2D}
\end{equation}
as such,
\begin{equation}
\nabla \cdot (\nabla \times \Psi) = 0 + 0 + \frac{\partial}{\partial z} \bigg(\frac{\partial \Psi_y}{\partial x} - \frac{\partial \Psi_x}{\partial y}\bigg) = 0.
\end{equation}


\item 

\tdplotsetmaincoords{70}{110}
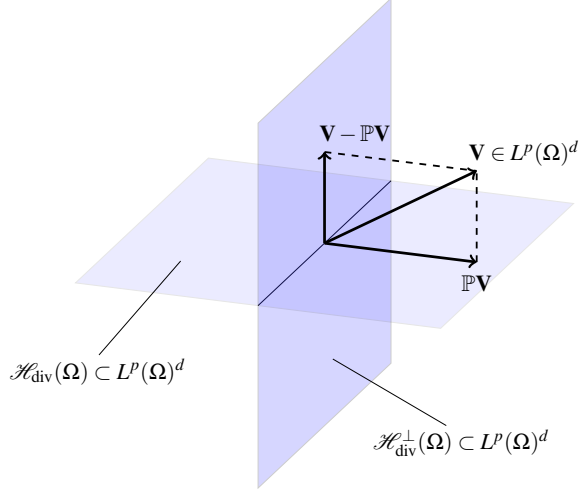
\begin{figure}
\centering
\noindent\resizebox{0.45\textwidth}{!}{
\begin{tikzpicture}[tdplot_main_coords]
\draw[-] (-3, 0, 0) -- (3, 0, 0);
\draw[fill=blue!80,opacity=0.2] (-3,0,-3) -- (-3,0,3) -- (3,0,3) -- (3,0,-3) -- cycle;
\draw[fill=blue!80,opacity=0.1] (-3,-3,0) -- (-3,3,0) -- (3,3,0) -- (3,-3,0) -- cycle;
\draw[very thick, ->](0,0,0)--(0,0,1.5);
\node[] at (0, 0.5, 1.85) {$\mathbf{V} - \mathbb{P}\mathbf{V}$};
\draw[very thick, ->](0,0,0)--(0,2.5,0);
\node[] at (0,2.5, -0.35) {$\mathbb{P}\mathbf{V}$};
\draw[very thick, ->](0,0,0)--(0,2.5,1.5);
\node[] at (-0.5, 3.1, 1.7) {$\mathbf{V} \in L^p(\Omega)^d$};
\draw[thick, dashed](0,0,1.5)--(0,2.5,1.5);
\draw[thick, dashed](0,2.5,0)--(0,2.5,1.5);
\node[] at (2, -3,-1.8) {$\mathcal{H}_{\text{div}}(\Omega) \subset L^p(\Omega)^d$};
\draw[-, very thin] (1, -2,-0.2) -- (4, -2.25, -0.75);
\node[] at (0, 2.25, -3) {$\mathcal{H}^\bot_{\text{div}}(\Omega) \subset L^p(\Omega)^d$};
\draw[-, very thin] (0, 2, -2.75) -- (1,  0.5,  -1.5);
\end{tikzpicture}}
\caption{Decompose a vector into divergence-free part and its orthogonal complement.}
\label{fig: hhd}
\end{figure}

Starting from sufficiently smooth vectors in $L^p(\Omega)^d$, the Helmholtz-Hodge decomposition theorem \cite{maria2003hhd} ensures the existence of its divergence-free part.
\begin{app_theorem}[Helmholtz-Hodge Decomposition (HHD)] A sufficiently smooth vector field $\mathbf{V} \in L^p(\Omega)^d$ on a domain $\Omega \subset \mathbb{R}^d$ with smooth boundary $\partial \Omega$, can be uniquely decomposed into a pure gradient field and a divergence-free vector field
\label{thm: hhd}
\begin{equation}
\mathbf{V} = \nabla p + \mathbf{u}, \quad p \in L^p(\Omega),\, \mathbf{u} \in \mathcal{H}_{{\text{div}}}(\Omega).
\end{equation}
\end{app_theorem} 

Theorem \ref{thm: hhd} can be interpreted as a mapping $\mathbb{P}$, which uniquely projects the smooth vector field $\mathbf{V}$ to its divergence-free part $\mathbf{u}$ by $\mathbb{P}\mathbf{V} = \mathbf{u}$, and representing $\mathbf{V}$ as $\mathbf{u}$ plus its orthogonal complement ($\nabla p$) (Fig. \ref{fig: hhd}). Therefore, the following conclusion is natural:
\begin{app_corollary}
$\mathbb{P}$ is an identity mapping on $H_{\text{div}}(\Omega)$. I.e, for $\forall \,\mathbf{V} \in \mathcal{H}_{{\text{div}}}(\Omega)$, $\mathbb{P} \mathbf{V} = \mathbf{V}$.
\label{thm: equi_div_free}
\end{app_corollary}

Further, for the special cases $\Omega \in \mathbb{R}^d$, $d=2,\, 3$, $\mathcal{H}_{\text{div}}(\Omega)$ is equivalent to the space of the curl of potentials with vanishing boundary condition \cite{dubois1990div_free}:
\begin{equation}
 \mathcal{H}_{\text{div}}(\Omega) \equiv \{\Psi \in  \mathcal{H}_{\text{curl}}(\Omega) \, \big| \, \Psi \cdot \mathbf{n}\big|_{\partial \Omega} = 0,\, \Omega \in \mathbb{R}^d \}\,.
\end{equation} 

Combined with Corollary \ref{thm: equi_div_free}, we have for $\forall\, \mathbf{V}\in L^p(\Omega)^d$, if $ \mathbf{V}$ is divergence-free, there exists a potential $\Psi \in \mathcal{H}_{\text{curl}}$ with vanishing boundary condition, which satisfies $\mathbf{V} = \nabla \times \Psi$.

\end{itemize}
\end{proof}

\clearpage
\vspace{0.5cm}
\section{Theorem 2: Symmetric PSD Tensor Representation by Spectral Decomposition}
\label{app: proof_psd}

{\it
\noindent For any tensor $\mathbf{D}  \in PSD(n)$, there exists an upper triangular matrix with zero diagonal entries, $\mathbf{B} \in \mathbb{R}^{\frac{n(n-1)}{2}}$, and a diagonal matrix with non-negative diagonal entries, $\boldsymbol{\Lambda} \in SD(n)$, satisfying:
\begin{equation}
\mathbf{D} = \mathbf{U}\, \boldsymbol{\Lambda} \, \mathbf{U}^T,\quad \mathbf{U} = exp(\mathbf{B} - \mathbf{B}^T) \in SO(n).
\label{app_eq: repre_psd}
\end{equation}
Conversely, for any upper triangular matrix with zero diagonal entries, $\mathbf{B} \in \mathbb{R}^{\frac{n(n-1)}{2}}$, and any diagonal matrix with non-negative diagonal entries, $\boldsymbol{\Lambda} \in SD(n)$, Eq.~(\ref{app_eq: repre_psd}) computes a symmetric PSD tensor, $\mathbf{D}\in PSD(n)$. 
}

\begin{proof}
First of all, let us define the following three special groups of interest:
\begin{app_definition}[The $n\times n$ symmetric PSD tensor group]
\begin{equation}
PSD(n) \equiv \{ \mathbf{D} \in \mathbb{R}^{n\times n}\,\big|\, \forall \mathbf{x}\in\mathbb{R}^n: \mathbf{x}^T\mathbf{D} \mathbf{x} \geq 0 \}.
\label{app: PSD}  
\end{equation}
\end{app_definition}

\begin{app_definition}[The special group of real orthogonal matrices]
\begin{equation}
SO(n) \equiv \{  \mathbf{U} \in \mathbb{R}^{n \times n}\, \big| \, \mathbf{U}^T\mathbf{U} = \mathbf{I}, \, \text{det}(\mathbf{U}) = 1 \}. 
\label{app: SO}
\end{equation}
\end{app_definition}

\begin{app_definition}[The special group of real diagonal matrices with all non-negative entries]
\begin{equation}
SD(n) \equiv \{diag(\lambda_1,\,...\, \lambda_n) \in \mathbb{R}^{n \times n} \,\big| \, \lambda_1,\,...,\, \lambda_n \geq 0\}.
\label{app: SD}
\end{equation}
\end{app_definition}

It is natural to consider the spectral decomposition form for a tensor field $\mathbf{D}\in PSD(n)$:  
\begin{app_theorem}[Spectral Decomposition for Symmetric Positive Semi-definite Matrix]
Let $\mathbf{D}$ be a $n \times n$ real symmetric matrix, which is positive semi-definite. Then $\mathbf{D}$ can be factorized as:
\begin{equation}
\mathbf{D} = \mathbf{U}\boldsymbol{\Lambda} \mathbf{U}^{T}, \quad \mathbf{U} \in SO(n),\, \boldsymbol{\Lambda}  \in SD(n),
\label{app: spectral}
\end{equation}
where columns of $\mathbf{U}$ are the eigenvectors of $\mathbf{D}$, the diagonal entries of $\boldsymbol{\Lambda}$ are the corresponding non-negative eigenvalues.
\label{app_thm: spectral_psd}
\end{app_theorem}

As such, any symmetric PSD tensor can be represented via its orthogonal eigenvectors and corresponding eigenvalues. Intuitively, we can represent a tensor $\mathbf{D}\in PSD(n)$ by representing its eigenvectors $\mathbf{U}\in SO(n)$ and eigenvalues $\boldsymbol{\Lambda} \in SD(n)$. 

\begin{itemize}
\item 
It is straightforward to represent a diagonal matrix $\boldsymbol{\Lambda} \in SD(n)$ with non-negative entries by simply imposing non-negativity in the implementation (For example via a \texttt{ReLU}). Representing an orthogonal matrix $\mathbf{U}$ needs more parametrization tricks.

In order to obtain an orthogonal matrix $\mathbf{U}$ by construction, we resort to the surjective Lie exponential mapping on $SO(n)$, which is in fact a compact and connected Lie group~\cite{lezcano2019spectral}. 
When seen as a sub-manifold of $\mathbb{R}^{n\times n}$ equipped with the metric induced from the ambient space $\langle \mathbf{X},\, \mathbf{Y} \rangle = tr(\mathbf{X}^T \mathbf{Y} )$, $SO(n)$ inherits a bi-invariant metric (i.e., the metric is invariant with respect to left and right multiplication by matrices of the group). The tangent space at the identity element of $SO(n)$ is a Lie algebra of $SO(n)$, which is essentially the group of skew-symmetric matrices, 
\begin{equation}
\mathfrak{so}(n) = \{  \mathbf{A} \in \mathbb{R}^{n \times n} \, \big| \,  \mathbf{A} +  \mathbf{A}^T = 0 \}.
\label{app: so}
\end{equation}
The Lie exponential map is a map from the Lie algebra $\mathfrak{g}$ of a Lie group $G$ to the group itself, which is an important tool for studying local structure of Lie groups. Specifically, it is written as:
\begin{align}
exp:\, \mathfrak{g} &  \to  G  \nonumber \\
\mathbf{A} & \mapsto exp(\mathbf{A}) := \mathbf{I} + \mathbf{A} + \frac{1}{2} \mathbf{A}^2 + \cdots.
\label{app: lie}
\end{align}
For the special group $SO(n)$, which is connected and compact~\cite{lezcano2019spectral}, the Lie exponential map from $\mathfrak{so}(n)$ to $SO(n)$ is surjective, which means $\forall \mathbf{U} \in SO(n)$, $\exists \mathbf{A} \in \mathfrak{so}(n), exp(\mathbf{A}) = \mathbf{U}$. Follow \cite{lezcano2019spectral}, an optimization problem from $SO(n)$ is equivalent to its corresponding optimization problem in Euclidean space, via Lie exponential mapping. In other words,  
\begin{equation}
\min\limits_{\mathbf{U}\in SO(n)}~f(\mathbf{U}) \quad   \iff  \quad \min\limits_{\mathbf{A}\in \mathfrak{so}(n)}~f(exp(\mathbf{A})).
\label{app: lie_opt}
\end{equation}
Furthermore, combining with the isomorphic mapping from vector space to $\mathfrak{so}(n)$~\cite{lezcano2019trivializations},
\begin{align}
\alpha:\, \mathbb{R}^{\frac{n(n-1)}{2}} &  \to  \mathfrak{so}(n) \nonumber \\
\mathbf{B} & \mapsto \mathbf{B}  - \mathbf{B}^T,
\label{app: iso}
\end{align}
where $\mathbb{R}^{\frac{n(n-1)}{2}}$ refers to the space of upper triangular matrices with zero diagonal entries, we reach the following equivalent optimization problem on the Euclidean space $\mathbb{R}^{\frac{n(n-1)}{2}}$.
\begin{equation}
\min\limits_{\mathbf{U}\in SO(n)}~f(\mathbf{U})  
\Leftrightarrow 
\min\limits_{\mathbf{A}\in \mathfrak{so}(n)}~f(exp(\mathbf{A}))
\Leftrightarrow 
\min\limits_{\mathbf{B}\in \mathbb{R}^{\frac{n(n-1)}{2}}}~f(exp(\mathbf{B - B}^T)).  
\end{equation}
In implementation, solving the exponential mapping is computationally expensive, especially for large scale matrices. Therefore, we use the {\it Cayley} retraction, a first order approximation for the exponential mapping~\cite{lezcano2019trivializations}:
\begin{equation}
exp(\mathbf{A}) \approx \phi(\mathbf{A}) = (\mathbf{I} + \frac{1}{2} \mathbf{A} ) (\mathbf{I} - \frac{1}{2} \mathbf{A}) ^{-1}.
\label{app: cayley}
\end{equation} 

Therefore, given any tensor $\mathbf{D} \in PSD(n)$, according to the existence of its spectral decomposition via Eq.~(\ref{app: spectral}) and the two surjective mappings of Eqs.~(\ref{app: lie}, \ref{app: iso}), there exists an upper triangular matrix with zero diagonal entries, $\mathbf{B} \in \mathbb{R}^{\frac{n(n-1)}{2}}$, and a diagonal matrix with non-negative diagonal entries, $\boldsymbol{\Lambda} \in SD(n)$, that satisfy Eq.~(\ref{app_eq: repre_psd}).

\item 
Conversely, given any upper triangular matrix with zero diagonal entries, $\mathbf{B} \in \mathbb{R}^{\frac{n(n-1)}{2}}$, and any diagonal matrix with non-negative diagonal entries, $\boldsymbol{\Lambda} \in SD(n)$, Eq.~(\ref{app_eq: repre_psd}) computes a symmetric PSD tensor, $\mathbf{D}\in PSD(n)$. Here, we give brief illustrations for the cases of $n=2,\, 3$, corresponding to diffusion tensors on 2D and 3D domains in our work respectively.

\begin{itemize}
\item {\it 2D tensors}. 
Denote $\mathbf{B} = 
	\begin{bmatrix}
	0 & s \\
	0 & 0
	\end{bmatrix}\in \mathbb{R}^{\frac{2(2-1)}{2}}$, applying Eq.~(\ref{app: iso}) we have
\begin{equation}
\mathbf{B} \mapsto \mathbf{A} := \alpha(\mathbf{B}) = 
	\begin{bmatrix}
	0 & s \\
	-s & 0
	\end{bmatrix} \in \mathfrak{so}(2), 
\end{equation}
applying Cayley retraction in Eq.~(\ref{app: cayley}) we have
\begin{equation}
\mathbf{A} \mapsto \mathbf{U} : = exp(\mathbf{A}) \approx 
	\begin{bmatrix} 
	1& \frac{s}{2} \\
	-\frac{s}{2}  & 1
	\end{bmatrix} \,
	\begin{bmatrix} 
	1& -\frac{s}{2} \\
	\frac{s}{2}  & 1
	\end{bmatrix} ^{-1}
  	= \frac{1}{s^2 + 4}\,
	\begin{bmatrix} 
	4-s^2& 4s \\
	-4s & 4-s^2
	\end{bmatrix} \in SO(2), 
	\label{app: 2d_psd}
\end{equation}
Combining with any given diagonal matrix with non-negative diagonal entries, $\boldsymbol{\Lambda} \in SD(2)$, we obtain $\mathbf{D} = \mathbf{U}\boldsymbol{\Lambda} \mathbf{U}^{T} \in PSD(2)$ by definition.

\item {\it 3D tensors}. 
Denote $\mathbf{B} = \begin{bmatrix}
	0 & s_1 & s_2 \\
	0 & 0 & s_3 \\
	0 & 0 & 0
	\end{bmatrix}\in \mathbb{R}^{\frac{3(3-1)}{2}}$, applying Eq.~(\ref{app: iso}) we have
\begin{equation}
\mathbf{B} \mapsto \mathbf{A} := \alpha(\mathbf{B}) = \begin{bmatrix}
	0 & s_1 & s_2 \\
	-s_1 & 0 & s_3 \\
	-s_2 & -s_3 & 0
	\end{bmatrix} \in \mathfrak{so}(3), 
\end{equation}
applying Cayley retraction in Eq.~(\ref{app: cayley}) we have
\begin{align}
\mathbf{A} \mapsto \mathbf{U} :&= exp(\mathbf{A})  \approx 
	\begin{bmatrix}
	1 & \frac{s_1}{2} &  \frac{s_2}{2} \\[0.6ex]
	- \frac{s_1}{2} & 1 &  \frac{s_3}{2} \\[0.6ex]
	- \frac{s_2}{2} & - \frac{s_3}{2} & 1
	\end{bmatrix}
	\begin{bmatrix}
	1 & -\frac{s_1}{2} &  -\frac{s_2}{2} \\[0.6ex]
	\frac{s_1}{2} & 1 &  -\frac{s_3}{2} \\[0.6ex]
	\frac{s_2}{2} & \frac{s_3}{2} & 1
	\end{bmatrix}^{-1} \nonumber \\[0.6ex]
	& =
	 \frac{1}{s_1^2+s_2^2+s_3^2+4}\,
	\begin{bmatrix}
	s_1^2  + s_2^2 + 4 & s_2s_3  & -s_1s_3 \\[0.6ex]
	s_2s_3 & s_1^2  + s_3^2 + 4 & s_1s_2 \\[0.6ex]
	-s_1s_3  & s_1s_2 & s_2^2  + s_3^2 + 4
	\end{bmatrix} \in SO(3), 
	\label{app: 3d_psd}
\end{align}

Combining with any given diagonal matrix with non-negative diagonal entries, $\boldsymbol{\Lambda} \in SD(3)$, we obtain $\mathbf{D} = \mathbf{U}\boldsymbol{\Lambda} \mathbf{U}^{T} \in PSD(3)$ by definition.

\end{itemize}
\end{itemize}
\vspace{-0.2cm}
\end{proof}


\clearpage

\section{PyTorch Advection-Diffusion PDE Toolkit: Brief Manual on Numerical Derivations}
\label{app: numericals}
Our advection-diffusion PDE toolkit is designed to solve, separately or jointly, advection and diffusion PDEs (1/2/3D). 
\begin{align}
\frac{\partial C({\mathbf{x}}, t)}{\partial t} =  \frac{\partial C({\mathbf{x}}, t)}{\partial t}\bigg |_{\text{adv}} +  \frac{\partial C({\mathbf{x}}, t)}{\partial t}\bigg |_{\text{diff}}  &= \underbrace{-  \nabla \left({\mathbf{V}}({\mathbf{x}}) \cdot C({\mathbf{x}}, t) \right)}_{\text{Fluid flow}} + \underbrace{\nabla \cdot \left({\mathbf{D}}({\mathbf{x}})\, \nabla C({\mathbf{x}}, t)\right)}_{\text{Diffusion}} \nonumber \\
(\nabla \cdot \mathbf{V} = 0) & = \underbrace{- {\mathbf{V}}({\mathbf{x}})\cdot\nabla C({\mathbf{x}}, t)}_{\text{Incompressible flow}} + \underbrace{\nabla \cdot \left({\mathbf{D}}({\mathbf{x}})\, \nabla C({\mathbf{x}}, t)\right)}_{\text{Diffusion}}.
\label{app_eq: adv_diff}  
\end{align}
One can choose to model the velocity field as a constant, a vector field, or a divergence-free vector field (for impressible flow). Furthermore, the diffusion field can be modeled as a constant, a non-negative scalar field, or a symmetric positive semi-definite (PSD) tensor field. The toolkit is implemented as a custom \texttt{torch.nn.Module} subclass, such that one can directly use it as an advection-diffusion PDE solver for data simulation or easily wrap it into DNNs or numerical optimization frameworks for inverse PDE problems, i.e., parameters estimation.


\subsection{Computing Advection}
\label{sec: impl_adv}
Given a certain velocity field $\mathbf{V}$, we have the advection equation generally written as:
\begin{equation}
 \frac{\partial C}{\partial t}\bigg |_{\text{adv}} = - \nabla \cdot ({\mathbf{V}}\, C),\quad \mathbf{V}\in\mathbb{R}^{d}(\Omega),
 \label{app: adv}
\end{equation}
where $d$ refers to the dimension of the domain $\Omega$, $C = C(\mathbf{x},\,t)$ denotes the mass concentration at location $\mathbf{x}\in\Omega$ and time $t$. For 1D domains, or the special case where the velocity is constant ($V$) over space, Eq.~(\ref{app: adv}) is simplified as $ \frac{\partial C}{\partial t}\bigg |_{\text{adv}} = - V \cdot \nabla C$.

\paragraph{Advection with Incompressible Flow}
\label{impl: div_free_deriv}
Assume the the velocity field $\mathbf{V}$ is divergence free, i.e., $\nabla \cdot {\mathbf{V}} = 0$, we have
\begin{equation}
 \frac{\partial C}{\partial t}\bigg |_{\text{adv}} = - \nabla \cdot ({\mathbf{V}}\, C) 
= - \nabla \cdot {\mathbf{V}}\, C - {\mathbf{V}} \cdot \nabla C 
= -  {\mathbf{V}} \cdot \nabla C. 
\label{eq: impl_piano_model}
\end{equation} 


\paragraph{First-order Upwind Scheme}
\label{sec: impl_upwind}

Given a 3D volumetric concentration image $C = \left(C_{i,j,k}\right)_{N_x \times N_y \times N_z }$ with grid sizes $\Delta x,\, \Delta y,\, \Delta z$, we approximate the partial differential derivatives in the advection equation (the right hand side of Eq.~(\ref{app: adv})) using upwind differences. We apply a first-order upwind scheme along each direction $x, \, y, \, z$, based on the corresponding velocity components $V^x, \, V^y, \, V^z$ of ${\mathbf{V}}$. Specifically, $\, \frac{\partial C}{\partial x}$ at $(i,\,j,\,k)$ is determined as:

\begin{equation}
\frac{\partial C}{\partial x}\bigg\rvert_{i, j, k} = 
    \begin{cases}
      \frac{C_{i,j,k} - C_{i-1,j,k}}{\Delta x}, & V^x_{i,j,k} \geq 0 \\[1ex]
      \frac{C_{i+1,j,k} - C_{i,j,k}}{\Delta x}, & V^x_{i,j,k} < 0
    \end{cases}, \quad
\frac{\partial C}{\partial y}\bigg\rvert_{i, j, k} = 
    \begin{cases}
      \frac{C_{i,j,k} - C_{i,j-1,k}}{\Delta y}, & V^y_{i,j,k} \geq 0 \\[1ex]
      \frac{C_{i,j+1,k} - C_{i,j,k}}{\Delta y}, & V^y_{i,j,k} < 0
    \end{cases},    
    \quad  
\frac{\partial C}{\partial z}\bigg\rvert_{i, j, k} = 
    \begin{cases}
      \frac{C_{i,j,k} - C_{i,j,k-1}}{\Delta z}, & V^z_{i,j,k} \geq 0 \\[1ex]
      \frac{C_{i,j,k+1} - C_{i,j,k}}{\Delta z}, & V^z_{i,j,k} < 0
    \end{cases}.
\end{equation}


\paragraph{Courant-Friedrichs-Lewy Condition for Advection}
\label{app: stab_adv}
To ensure that the above upwind scheme is numerically stable when integrating Eq.~(\ref{app: adv}), the following Courant-Friedrichs-Lewy condition (CFL) should be satisfied:
 \begin{equation}
 c = \sum_{ax\in\{x,y,z\}} \frac{V^{ax}\, \Delta t}{\Delta ax} \leq c_{\text{max}},
 \label{eq: cfl_adv}
 \end{equation}
 where $c_{\text{max}}$ approximately equals to 1 when applying explicit methods for ordinary differential equations (ODEs) \cite{gottlieb2000ssp,leveque2002}.


\subsection{Computing Diffusion}
\label{sec: impl_diff}

Given a certain diffusion field $\mathbf{D}$, the diffusion equation is written as:
\begin{equation}
 \frac{\partial C}{\partial t}\bigg |_{\text{diff}} =  \nabla \cdot ({\mathbf{D}}\, \nabla C),\quad \mathbf{D}\in\mathbb{R}^{n\times n}(\Omega).
 \label{app: diff}
\end{equation}
where $C = C(\mathbf{x},\,t)$ denotes the mass concentration at location $\mathbf{x}\in\Omega$ and time $t$. For 1D domains, or the special case where the diffusion is a constant or a scalar field ($D$), Eq.~(\ref{app: diff}) can be simplified to $ \frac{\partial C}{\partial t}\bigg |_{\text{{diff}}} =  D \cdot \Delta C$.

\paragraph{Nested Central-Forward-Backward Differences}
\label{sec: impl_nested}
In order to obtain a more stable numerical discretization scheme for the diffusion part in Eq.~(\ref{app: diff}), in practice, we explicitly compute its expanded formula. Here, we give the brief illustration of our discretization scheme for $n=2,\, n=3$, corresponding to the diffusion tensors on 2D and 3D domains, respectively. ($\mathbf{D}$ are represented via Eq.~(\ref{app: 2d_psd}) (2D) or Eq.~(\ref{app: 3d_psd}) (3D) when the diffusion fields are assumed to be symmetric PSD.) 

\begin{itemize}
\item {\it 2D tensors.} For symmetric PSD tensor field $\mathbf{D} = \begin{bmatrix}
	Dxx & Dxy \\ 
	Dxy & Dyy
	\end{bmatrix}\in \mathbb{R}^{2 \times 2}(\Omega)$:
\begin{align}
	\frac{\partial C}{\partial t}\bigg |_{\text{diff}}  & = \nabla \cdot \Bigg(\begin{bmatrix}
	Dxx & Dxy \\ 
	Dxy & Dyy
	\end{bmatrix}\cdot \begin{bmatrix}
	\frac{\partial C}{\partial x} \\
	\frac{\partial C}{\partial y}
	\end{bmatrix} \Bigg )  = \frac{\partial }{\partial x} \, \bigg (Dxx\, \frac{\partial C}{\partial x} + Dxy\, \frac{\partial C}{\partial y} \bigg) + \frac{\partial }{\partial y} \, \bigg (Dxy\, \frac{\partial C}{\partial x} + Dyy\, \frac{\partial C}{\partial y} \bigg) \nonumber \\[0.7ex]
	& = \underbrace{ \bigg ( \frac{\partial Dxx}{\partial x} \, \frac{\partial C}{\partial x}  + \frac{\partial Dxy}{\partial x} \, \frac{\partial C}{\partial y} + \frac{\partial Dxy}{\partial y} \, \frac{\partial C}{\partial x} + \frac{\partial Dyy}{\partial y} \, \frac{\partial C}{\partial y} \bigg )}_{\text{(a)}} 
	+ \underbrace{\bigg( Dxx \, \frac{\partial^2 C}{\partial x^2} + 2\,Dxy \, \frac{\partial^2 C}{\partial x\partial y} + Dyy \, \frac{\partial^2 C}{\partial y^2} \bigg)}_{\text{(b)}}.
	\end{align}
	
\item For symmetric positive semi-definite tensor $\mathbf{D} = \begin{bmatrix}
	Dxx & Dxy & Dxz \\ 
	Dxy & Dyy & Dyz \\
	Dxz & Dyz & Dzz
	\end{bmatrix}\in \mathbb{R}^{3 \times 3}(\Omega)$:
	
\begin{align}
	\frac{\partial C}{\partial t}\bigg |_{\text{diff}}  & = \nabla \cdot \Bigg(\begin{bmatrix}
	Dxx & Dxy & Dxz \\[0.7ex]
	Dxy & Dyy & Dyz \\[0.7ex]
	Dxz & Dyz & Dzz
	\end{bmatrix}\cdot \begin{bmatrix}
	\frac{\partial C}{\partial x} \\[0.7ex]
	\frac{\partial C}{\partial y} \\[0.7ex]
	\frac{\partial C}{\partial z}
	\end{bmatrix} \Bigg ) \nonumber \\[0.7ex]
	& = \frac{\partial }{\partial x} \, \bigg(Dxx\, \frac{\partial C}{\partial x} + Dxy\, \frac{\partial C}{\partial y} + Dxz\, \frac{\partial C}{\partial z} \bigg) 
	+ \frac{\partial }{\partial y} \, \bigg (Dxy\, \frac{\partial C}{\partial x} + Dyy\, \frac{\partial C}{\partial y} + Dyz\, \frac{\partial C}{\partial z} \bigg )  	
	+ \frac{\partial }{\partial z} \, \bigg (Dxz\, \frac{\partial C}{\partial x} + Dyz\, \frac{\partial C}{\partial y} + Dzz\, \frac{\partial C}{\partial z} \bigg) \nonumber \\[0.7ex]
	& =  \underbrace{ 
	  \Bigg (  \bigg (\frac{\partial Dxx}{\partial x} + \frac{\partial Dxy}{\partial y} + \frac{\partial Dxz}{\partial z} \bigg) \frac{\partial C}{\partial x}  
	 + \bigg (\frac{\partial Dxy}{\partial x} + \frac{\partial Dyy}{\partial y} + \frac{\partial Dyz}{\partial z} \bigg) \frac{\partial C}{\partial y}  
	 + \bigg (\frac{\partial Dxz}{\partial x} + \frac{\partial Dyz}{\partial y} + \frac{\partial Dzz}{\partial z} \bigg) \frac{\partial C}{\partial z} \Bigg )
	  }_{\text{(a)}} \nonumber \\
	& \qquad +  \underbrace{ 
	  \Bigg (  \bigg (\frac{\partial Dxx}{\partial x} + \frac{\partial Dxy}{\partial y} + \frac{\partial Dxz}{\partial z} \bigg) \frac{\partial C}{\partial x}  
	 + \bigg (\frac{\partial Dxy}{\partial x} + \frac{\partial Dyy}{\partial y} + \frac{\partial Dyz}{\partial z} \bigg) \frac{\partial C}{\partial y}  
	 + \bigg (\frac{\partial Dxz}{\partial x} + \frac{\partial Dyz}{\partial y} + \frac{\partial Dzz}{\partial z} \bigg) \frac{\partial C}{\partial z} \Bigg )
	  }_{\text{(a)}}  
\end{align} 

\end{itemize}

Specifically, we use the central difference scheme for discretizing all first-order spatial derivatives in (a), and nested forward-backward differences for all second-order operators in (b).



\paragraph{Mesh Fourier Number for Diffusion}
\label{app: stab_diff}
To ensure that the above finite difference scheme for diffusion is numerically stable when integrating forward in time, the following condition should be satisfied\footnote{See \url{https://hplgit.github.io/fdm-book/doc/pub/book/sphinx/._book011.html}.}:
\begin{equation}
F = \sum_{ax \in \{x, y, z\}} \frac{D^{ax} \, \Delta t}{\Delta ax^2} \leq \frac{1}{2},
\label{eq: stab_diff}
\end{equation}
 where $D^{ax},\, ax \in \{x,y,z\}$ refer to the diagonal entries $Dxx,\, Dyy,\, Dzz$ of the diffusion tensor $\mathbf{D}$.


\subsection{Numerical Solution}  
As introduced above, after discretizing all the spatial derivatives on the right side of Eq.~(\ref{app_eq: adv_diff}), we obtain a system of ordinary differential equations (ODEs), which can be solved by numerical integration. We then use the RK45 method to advance in time ($\delta t$) to predict $\widehat{C}^{t+\delta t}$. Note when the input mass transport time-series has relatively large temporal resolution ($\Delta t$), the chosen $\delta t$ should be smaller than $\Delta t$ to satisfy the stability conditions discussed in Supp.~\ref{app: stab_adv}-\ref{app: stab_diff}, thereby ensuring stable numerical integration. 

\clearpage 

\section{Brain Advection-Diffusion Dataset}
\label{app: ixi}

\input{sub/appendix/fig/v_generator}
\input{sub/appendix/fig/d_generator}
\input{sub/appendix/fig/movie_generator}

Our brain advection-diffusion simulation dataset is based on the public \texttt{IXI} brain dataset\footnote{Available for download: \url{http://brain-development.org/ixi-dataset/}.}, from which we use 200 patients with complete collections of T1-/T2-weighted images, 
magnetic resonance angiography (MRA) image, and diffusion-weighted images (DWI) with 15 directions for the simulation of 3D divergence-free velocity vector and symmetric PSD diffusion tensor fields. All above images are resampled to isotropic spacing ($1\,mm$), rigidly registered intra-subject, and skull-stripped using \texttt{ITK}\footnote{Code in \url{https://github.com/InsightSoftwareConsortium/ITK}.}. 


\subsection{Divergence-free Velocity Vector Field Simulation (Fig.~\ref{app_fig: v_gen})}
\label{app_sec: velocity}

\paragraph{Blood Vessel Segmentation}
Brain blood vessels are segmented by \texttt{ITK-TubeTK} using T1-/T2-weighted and MRA images, where T1-/T2-weighted images are used to obtain more robust segmentation results\footnote{Code in \url{https://github.com/InsightSoftwareConsortium/ITKTubeTK/tree/master/examples/MRA-Head}.}.

\paragraph{Principal Vessel Direction Inference} 
We infer a vessel's directions by analyzing the spectral decomposition of the second order derivatives (Hessian, $\mathcal{H}_I$) of the vessel segmentation image $I$:
\begin{equation}
\mathcal{H}_I = \mathbf{U}_I\boldsymbol{\Lambda}_I \mathbf{U}_I^{T} = 
	\begin{bmatrix}
	\mathbf{u}_1 &  \mathbf{u}_2 & \mathbf{u}_3 
	\end{bmatrix}
	\begin{bmatrix}
	\lambda_1 &  & \\ \vspace{1mm} 
	 & \lambda_2 &  \\ \vspace{1mm} 
	 & & \lambda_3
	\end{bmatrix}
	 \begin{bmatrix} 
	 \mathbf{u}_1^T \\ \vspace{1mm} \mathbf{u}_2^T \\ \vspace{1mm}  \mathbf{u}_3^T 
	 \end{bmatrix}.
\end{equation}
where the eigenvalues are sorted by their absolute values as $|\lambda_1| \leq |\lambda_2| \leq |\lambda_3|$. Following Frangi et al.~\cite{frangi1998vessel}, the eigenvector $\mathbf{u}_1$ corresponding to $\lambda_1$ with the smallest absolute value, indicates the direction along the vessel, while the remaining eigenvectors $\mathbf{u}_2,\, \mathbf{u}_3$ form a base for the orthogonal plane crossing the vessel. We therefore take $\mathbf{u}_1$ as the vessel direction maps.

In order to reduce the effect from the sign ambiguity of eigenvector directions and thus obtain more consistent flow velocity directions, we first extract the centerline of the vessel map and determine the flow directions along the centerline. Specifically, we start at the point with the largest MRA intensity along the centerline, from which we set the flow direction to its nearest centerline point. Likewise, we traverse all points along the centerline and successively assign directions for the nearest point along the centerline. Afterwards, velocity directions of the remaining voxels within the vessels are set to be the same as its nearest centerline point. I.e., flip the sign of the velocity component if it is not the same with that of its nearest centerline point. In this way, we ensure a velocity field with consistent flow directions.

\paragraph{Divergence-free Velocity Simulation} 
The velocity field $\mathbf{V}$ is obtained, by multiplying the vessel direction maps with the MRA intensities, to reflect the local velocity magnitudes. We scale the value range of the velocity field to $[-1,\, 1]$, and additionally compute velocity fields with $50\%$ of the original velocity values. 
 
Divergence-free velocity fields $\mathbf{V}_{\text{div}}$ are then estimated from $\mathbf{V}$ via Eq.~(\ref{app_eq: repre_v}) using numerical optimization, by encouraging $\mathbf{V}_{\text{div}} := \nabla \times \boldsymbol{\Psi}$ close to $\mathbf{V}$:
\begin{equation}
\min\limits_{\boldsymbol{\Psi}}\, \big\| \nabla \times \boldsymbol{\Psi} - \mathbf{V} \big\|_2^2, \quad \boldsymbol{\Psi} \cdot \mathbf{n}\big|_{\partial \Omega} = 0,\, \boldsymbol{\Psi} \in \mathbb{R}^3(\Omega).
\label{app_eq: opt_adv}
\end{equation}


\subsection{Symmetric PSD Diffusion Tensor Field Simulation (Fig.~\ref{app_fig: d_gen})}
\label{app_sec: diffusion}

\paragraph{Diffusion Tensor Estimation}
Diffusion tensors are reconstructed from the DWIs using \texttt{Dipy}\footnote{Code in \url{https://github.com/dipy/dipy}.}~\cite{dipy}. Specifically, the diffusion tensors are estimated based on the following equation:
\begin{equation}
\frac{S(\mathbf{g}, \, b)}{S_0} = e^{-b\mathbf{g}^T\mathbf{D}\mathbf{g}},
\label{eq: dti}
\end{equation}
where $\mathbf{g}$ is a unit vector indicating the direction of measurement and $b$ are the parameters of measurement, e.g., incorporating the strength and duration of the diffusion-weighting gradient. $S(\mathbf{g}, \, b)$ is the diffusion-weighted signal measured and $S_0$ is the signal measured with no diffusion weighting.  $\mathbf{D}$ is the symmetric PSD tensor which contains six free parameters to be fit:
\begin{equation}
\mathbf{D} = \begin{bmatrix}
D_{xx} & D_{xy} & D_{xz} \\
D_{xy} & D_{yy} & D_{yz} \\
D_{xz} & D_{yz} & D_{zz} 
\end{bmatrix}\,.
\end{equation}

We scale the value range of the diffusion tensor fields for each subject sample to $[-0.2,\, 0.2]$, and additionally compute diffusion fields with $50\%$ of the original diffusion values.

\paragraph{Diffusion Scalar Maps} 
Upon obtaining the diffusion tensor $\mathbf{D}$, we decompose it into eigenvectors ($\mathbf{U}$) and eigenvalues ($\boldsymbol{\Lambda}$):
\begin{equation} 
\mathbf{D} = \mathbf{U}\boldsymbol{\Lambda} \mathbf{U}^{T} = 
	\begin{bmatrix}
	\mathbf{u}_1 &  \mathbf{u}_2 & \mathbf{u}_3 
	\end{bmatrix}
	\begin{bmatrix}
	\lambda_1 &  & \\ \vspace{1mm} 
	 & \lambda_2 &  \\ \vspace{1mm} 
	 & & \lambda_3
	\end{bmatrix}
	 \begin{bmatrix} 
	 \mathbf{u}_1^T \\ \vspace{1mm} \mathbf{u}_2^T \\ \vspace{1mm}  \mathbf{u}_3^T ,
	 \end{bmatrix}.
\end{equation}
from which we can further visualize the diffusion scalar maps~\cite{mukherjee2008dti}, which describe tensor structure. In this work, we consider (1) Trace ($tr_{\mathbf{D}}$),
\begin{equation}
tr_{\mathbf{D}} = \lambda_1 + \lambda_2 + \lambda_3;
\end{equation}
 (2) Fractional anisotropy (FA),
 \begin{equation}
\text{FA} =  \sqrt{\frac{1}{2}}\sqrt{\frac{(\lambda_1 - \lambda_2)^2 + (\lambda_2 - \lambda_3)^2 + (\lambda_3 - \lambda_1)^2}{\lambda_1^2 + \lambda_2^2 + \lambda_3^2}};
\end{equation} 
(3) Principal diffusion orientation ($\mathbf{U}_{\text{prin}}$), the eigenvector $\mathbf{U}_{\text{prin}} = [u_x,\, u_y,\, u_z]^T$ corresponding to the largest eigenvalue.

Further, we also visualize the color-by-orientation (CbO) map, by assigning colors to voxels based on a combination of FA and $\mathbf{U}_{\text{prin}}$, i.e.,
\begin{equation}
{\text{Red}} = {\text{FA}} \cdot u_x;\,\, {\text{Green}} = {\text{FA}} \cdot u_y;\,\, {\text{Blue}} = {\text{FA}} \cdot u_z.
\end{equation}

 
\subsection{Brain Advection-diffusion Time-series Simulation (Fig.~\ref{app_fig: movie_gen})} 
\label{app_sec: movie}

For each brain advection-diffusion sample, the initial concentration state is assumed to be given by the MRA image with intensity ranges rescaled to $[0,\,1]$.
Time-series (length $N_T = 40$, interval $\Delta t = 0.1\,s$) are then simulated given the computed divergence-free velocity fields (Supp.~\ref{app_sec: velocity}) and symmetric PSD diffusion tensor fields (Supp.~\ref{app_sec: diffusion}) by our advection-diffusion PDE solver (Supp.~\ref{app: numericals}). Thus the simulated dataset includes 800 brain advection-diffusion time-series (4 time-series for each of the 200 subjects, based on the four combinations of the simulated two velocity fields and two diffusion fields).
 





\end{document}